\documentclass[preprint,floatfix,longbibliography,superscriptaddress,citeautoscript,nobibnotes,prb]{revtex4-1}
\usepackage{amsmath}
\usepackage{amssymb}
\usepackage{graphicx}
\usepackage{bm}
\usepackage{verbatim}
\usepackage{color}
\usepackage{subfigure}
\usepackage{natbib}
\usepackage{braket}
\usepackage[english]{babel}
\usepackage{blindtext}
\usepackage[pdftex, breaklinks,colorlinks, citecolor=blue, urlcolor=blue]{hyperref}

\usepackage[labelfont=bf,figurename=Fig.]{caption} 

\usepackage{lineno}

\begin{document}

\title{Nonclassical energy squeezing of a macroscopic mechanical oscillator}
\author{X. Ma}
\email[direct correspondence to ]{xizheng.ma@colorado.edu}\affiliation{JILA, National Institute of Standards and Technology and University of Colorado, and Department of Physics, University of Colorado, Boulder, Colorado 80309, USA}

\author{J. J. Viennot}
\affiliation{JILA, National Institute of Standards and Technology and University of Colorado, and Department of Physics, University of Colorado, Boulder, Colorado 80309, USA}
\affiliation{Univ. Grenoble Alpes, CNRS, Institut N\'eel, 38000 Grenoble, France 
}

\author{S. Kotler}
\affiliation{Department of Physics, University of Colorado, Boulder, CO 80309, USA.}
\affiliation{National Institute of
Standards and Technology (NIST), Boulder, Colorado 80305, USA.}

\author{J. D. Teufel}
\affiliation{National Institute of
Standards and Technology (NIST), Boulder, Colorado 80305, USA.}

\author{K. W. Lehnert}
\affiliation{JILA, National Institute of Standards and Technology and University of Colorado, and Department of Physics, University of Colorado, Boulder, Colorado 80309, USA}
\maketitle

\textbf{Optomechanics and electromechanics have made it possible to prepare macroscopic mechanical oscillators in their quantum ground states\cite{Teufel2011GndStateCooling}, in quadrature squeezed states\cite{Wollman2015MechSqueezing,Delaney2019Squeezing}, and in entangled states of motion\cite{Palomaki2013Entanglement,Ockeloen-Korppi2018EntangleTwoMech}. 
In addition to coaxing ever larger and more tangible objects into a regime of quantum behavior, this new capability has encouraged ideas of using mechanical oscillators in the processing and communication of quantum information and as precision force sensors operating beyond the standard quantum limit\cite{Caves1980WeakClassicalForceMeasurement,Braginsky1980QNDelectromechanics}. 
But the effectively linear interaction between motion and light or electricity precludes access to the broader class of quantum states of motion, such as cat states or energy squeezed states. Indeed, early optomechanical proposals\cite{Thompson2008StrongDispersiveCoupling,Jayich2008QuadraticOptomechanics} noted the possibility to escape this restriction by creating strong quadratic coupling of motion to light. Although there have been experimental demonstrations of quadratically coupled optomechanical systems\cite{Thompson2008StrongDispersiveCoupling,Brawley2016XsquaredPreparationBimodalState,Leijssen2017QuadraticCoupling}, these have not yet accessed nonclassical states of motion. 
Here we create nonclassical states by quadratically coupling motion to the energy levels of a Cooper-pair box (CPB) qubit. 
By monitoring the qubit's transition frequency, we detect the oscillator's phonon distribution rather than its position. 
Through microwave frequency drives that change both the state of the oscillator and qubit, we then dissipatively stabilize the oscillator in a state with a large mean phonon number of 43 and sub-Poissonian number fluctuations of approximately 3.  In this energy squeezed state we observe a striking feature of the quadratic coupling: the recoil of the mechanical oscillator caused by qubit transitions, closely analogous to the vibronic transitions in molecules\cite{Franck1925,Condon1926}. }

The ability to access a broad range of quantum states with mechanical oscillators has many applications and is an enduring ambition in the fields of opto- and electromechanics. As mechanical oscillators are linear at the quantum scale, arbitrary quantum control over them requires an extrinsic nonlinearity such as a nonlinear source\cite{Reed2017FaithfulConversion} or detector\cite{Riedinger2016HeraldMech,Lecocq2015VacuumFluctuationOptomechanics}. Alternatively, a mechanical oscillator can be coupled to an ancillary system in a nonlinear manner. For example, opto- and electromechanical systems routinely use the inherently nonlinear radiation pressure interaction between the motion of a mechanical oscillator and the energy of an ancillary optical or electrical cavity.

However, the radiation pressure interaction is intrinsically weak; thus, most experiments operate with a large cavity drive, increasing the coupling strength but yielding a linear interaction between oscillator motion and cavity field. In references \cite{Thompson2008StrongDispersiveCoupling,Jayich2008QuadraticOptomechanics}, the authors propose a solution: by coupling the square of the oscillator motion to the cavity energy, the drive-enhanced coupling remains nonlinear. Despite new theoretical insights \cite{Miao2009ResidualRadiationCoupling, Jacobs2009EngineeringControl,Dellantonio2018ElectromechG1G2Ratio,Hauer2018OptomechG1G2Ratio} and rapid experimental progress\cite{Doolin2014StationaryQuadraticOptomech,Paraiso2015ExpTunableQuadraticCoupling,Brawley2016XsquaredPreparationBimodalState,Leijssen2017QuadraticCoupling}, quadratic coupling schemes for opto- and electromechanics\cite{Braginsky1980QNDelectromechanics} have not yielded nonclassical states in mechanical oscillators.

\begin{figure} [h]
    \centering
    \includegraphics{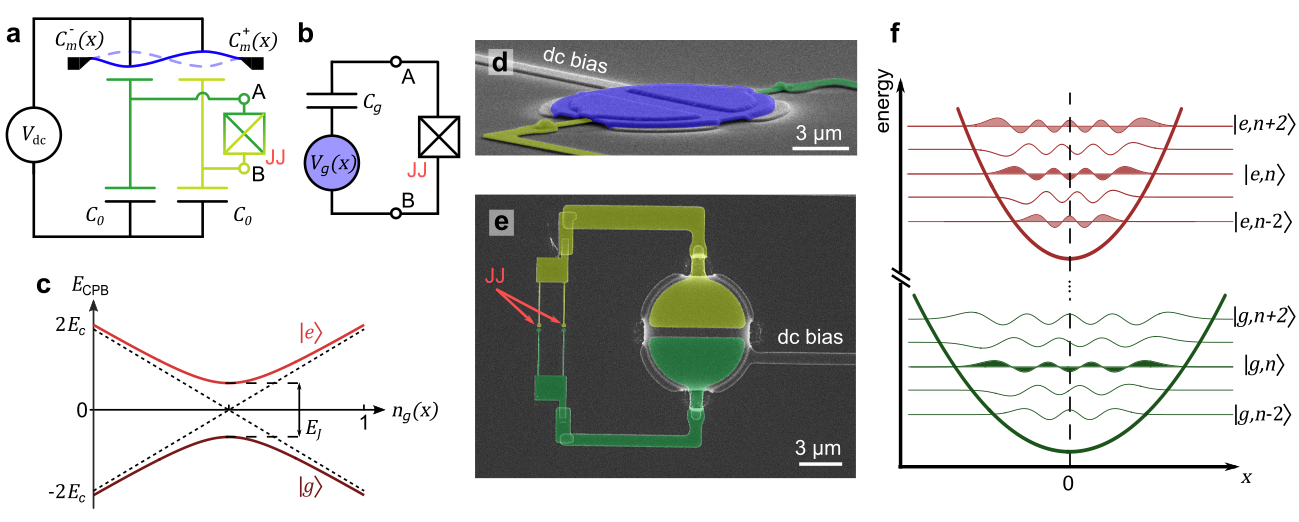}
    \caption{\textbf{Quadratically coupled electromechanics}}
    \label{fig:1}
\end{figure}

To overcome the weak coupling of electromechanics, in which the zero-point motion of the mechanical oscillator alters the tiny zero-point electrical energy stored in the capacitor of a resonant circuit, we arrange for motion to alter the large electrostatic energy stored in the capacitor of a CPB qubit by an applied dc voltage\cite{LaHaye2009NanomechanicalQubit,Pirkkalainen2013HybridResonator,Viennot2018Phonon} (Fig.~\ref{fig:1}a-b). At a point of charge degeneracy in the qubit, this arrangement creates a quadratic coupling between the oscillator position and qubit energy (Fig.~\ref{fig:1}c). We use this quadratic coupling to adiabatically stabilize a mechanical oscillator into an energy squeezed state with average phonon number of 43 and variance less than 11, yielding a ratio of $F = 0.257^{+0.002}_{-0.001}$ far below the classical limit of $F\geq 1 $\cite{Short1983ObserveSubPoissonian}. As a consequence of creating this high-energy number-squeezed state, we also observe sidebands in the qubit spectrum that reveal qubit excitation processes that create or annihilate phonons by pairs. 

To achieve strong quadratic interaction between the mechanical oscillator and a superconducting qubit, we embed a mechanically compliant elliptical disk into the microwave circuit shown in Fig.~\ref{fig:1}d-e. The mechanical oscillator is the anti-symmetric, second mode of the suspended disk with a resonant frequency $\omega_m \approx 2\pi \times 25$~MHz. Underneath the disk, two aluminum electrodes are placed at the anti-nodes of motion to form two mechanically compliant capacitors. The electrodes are connected through two Josephson junctions in parallel, creating a flux-tunable CPB qubit\cite{Bouchiat1998QuantumCoherencewithSingleCooperPair,Makhlin2000QuantumEngineeringWithJJdevices}. 
A static voltage $V_\text{dc}$ applied to the disk couples motion to the qubit energy as illustrated by an approximate electromechanical schematic in Fig.~\ref{fig:1}a (see supplementary material Sec.~IB).  The symmetry of the capacitor network is broken by the anti-symmetric motion of the oscillatory mode, yielding a CPB qubit with a gate voltage $V_g(x)$ proportional to the oscillator's coordinate (Fig.~\ref{fig:1}b). By applying $V_\text{dc} = 6$~V, we achieve a coupling rate of $g_m \approx 2\pi \times 22$~MHz, orders of magnitude larger than the values achieved using radiation pressure electromechanical coupling\cite{Teufel2011CircuitRegime}.
Operating at the charge degeneracy point, the qubit energy is first order insensitive to the oscillator's motion, but the second order quadratic coupling\cite{Miao2009ResidualRadiationCoupling} $2\chi_m \approx (2g_m)^2/(E_J/\hbar) \approx 2\pi \times 0.52$~MHz is large enough to profoundly affect the qubit and oscillator dynamics.  

Although it is expedient\cite{Viennot2018Phonon} to approximate a strong quadratic coupling of motion to a qubit using the dispersive limit of Jaynes-Cummings Hamiltonian, familiar from circuit quantum electrodynamics (cQED), this approximation fails to fully capture phenomena associated with the large separation in energy scales ($g_m, \omega_m \ll \omega_q$). Instead, the oscillator's position behaves as a slow coordinate\cite{Ashhab2010Qubit-oscillatorStatesUSC} moving in a potential modified by the state of the qubit\cite{Zueco2009BeyongRWA,Beaudoin2011DissipationUSC} (see also supplementary material sec.IIA),

\begin{equation} \label{Hamiltonian}
    H = \frac{1}{2}\hbar\omega_q \hat{\sigma}_z + \frac{\hat{p}^2}{2m}  + \frac{1}{2}k \bigg(1 + \frac{2 \chi_m }{\omega_m}\hat{\sigma}_z \bigg)  \hat{x}^2,
\end{equation}

\noindent where $\hat{x}$ and $\hat{p}$ are the position and momentum operators of the mechanical oscillator respectively, $m$ and $k$ are the mass and the spring constant, $\omega_m$ and $\omega_q$ are the bare mechanical and qubit frequencies, $\hat{\sigma}_z$ is the qubit Pauli operator, and
$\chi_m = g_m^2/(\omega_q-\omega_m) + g_m^2/(\omega_q+\omega_m) $ includes the Bloch-Siegert shift\cite{Bloch1940BSshift,Zueco2009BeyongRWA,Beaudoin2011DissipationUSC}. 
The slow mechanical oscillator thus experiences a sudden compression of its effective spring constant $k(\hat{\sigma}_z) = k(1+2\chi_m \hat{\sigma}_z/\omega_m)$ when the qubit changes state, which simultaneously alters the mechanical frequency $\omega_m(\hat{\sigma}_z) = \sqrt{k(\hat{\sigma}_z)/m}$ and the mechanical impedance $Z_m(\hat{\sigma}_z) = \sqrt{k(\hat{\sigma}_z)m}$. 
Because the impedance determines the spatial scale of the qubit-state-dependent mechanical wavefunctions\cite{Shankar}, wavefunctions that differ by even phonon numbers and with opposite qubit excitation are not orthogonal (Fig.~\ref{fig:1}f).
Consequently, the qubit spectrum will exhibit sideband features associated with the pair-wise creation and destruction of phonons. 
Although they are analogous to the sideband transitions in cQED systems\cite{Blais2007QuantumInformationProcessing,Wallraff2007cQEDSideband} where the oscillator and qubit frequencies are comparable, the small $\omega_m$ ($\ll \omega_q$) means these electromechanical sidebands not only appear close to the qubit frequency, but are also likely to be excited by a change in the qubit state. Indeed, a qubit transition is likely to alter the phonon number when the qubit-induced change in mean mechanical energy $\delta k \braket{\hat{x}^2}/2 = 2\hbar \chi_m n$ is larger than the energy of two phonons $2\hbar\omega_m$.
Thus, the condition $\chi_m n/\omega_m \gtrsim 1$ signifies the entry into a new regime where phonon-altering qubit transitions become dominant over phonon-preserving transitions.

Nevertheless, similar to cQED experiments\cite{Schuster2007PhotonNumberSplit}, the qubit-state-dependent mechanical frequency $\omega_m(\hat{\sigma}_z)$ leads to a phonon-number-dependent Stark shift on the qubit resonance. We use this shift in qubit frequency to determine the phonon distribution of the mechanical oscillator with a precision given by the phonon-number-sensitivity\cite{Viennot2018Phonon} $\xi = \Gamma_2^\star/2\chi_m \approx 7.1$ phonons, where $\Gamma_2^\star$ is the qubit decoherence rate.
The probability of exciting the qubit as a function of the frequency of a weak qubit-drive tone (qubit spectroscopy) is given by
\begin{equation}    \label{eq:DressedSpectrum}
    P_e(\omega) = \sum_n P(n) \times P_e^{\ket{n}}(\omega),
\end{equation}
a convolution between the phonon distribution $P(n)$ and the qubit spectrum with exactly $n$ phonons in the mechanical oscillator $P_e^{\ket{n}}(\omega)$. 
In contrast to reference\cite{Viennot2018Phonon} where $P_e^{\ket{n}}(\omega)$ are treated simply as Stark shifted Lorentzians, we employ a deconvolution\cite{Richardson1972LRreconstruction} procedure that accounts for the sideband transitions in these qubit spectra. This more accurate procedure is necessary because we endeavor to create states with large average phonon number and small distributions. 
However, its implementation requires that we accurately determine $P_e^{\ket{n}}(\omega)$. As explained in the supplementary material Sec.III, this determination is experimentally achieved by setting $V_\text{dc} = 0$~V and simulating the effect of motion with a classical ac-voltage that modulates the gate-charge at $\omega_m$.

\begin{figure} [h]
    \centering
    \includegraphics{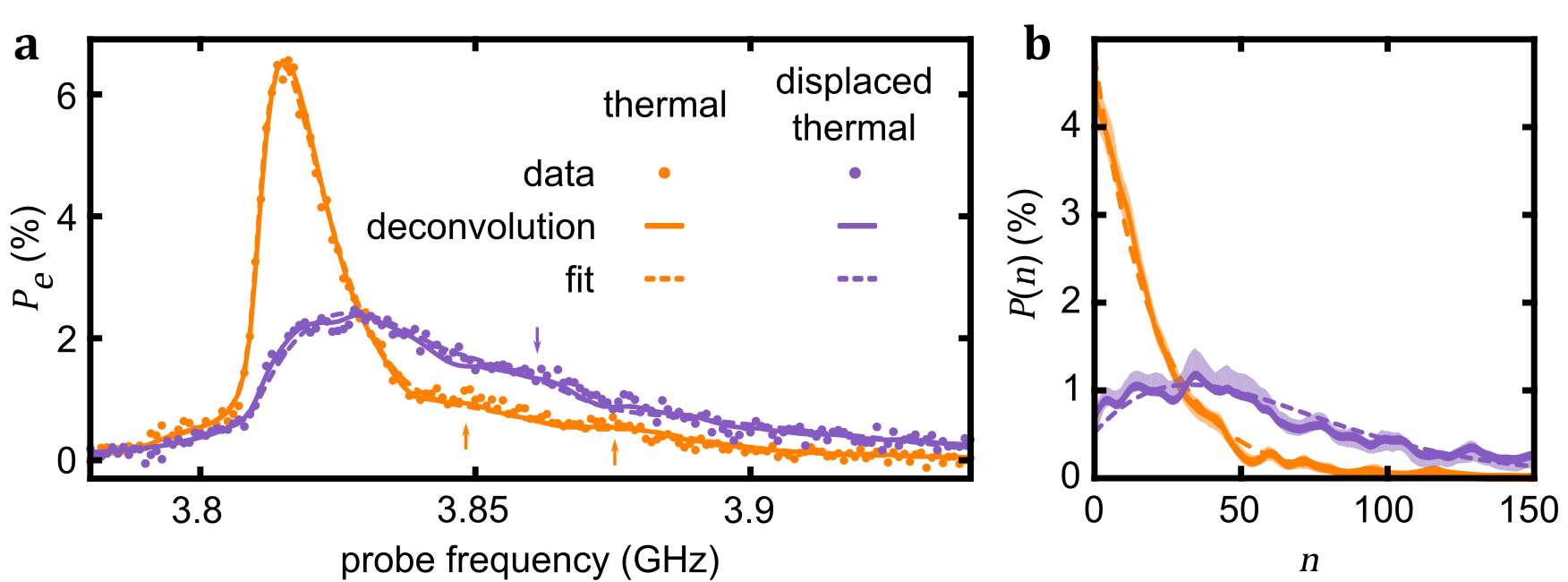}
    \caption{\textbf{Determining the phonon distribution from a qubit spectrum}}
    \label{fig:2}
\end{figure}

To validate the deconvolution procedure, we demonstrate it on thermal and displaced-thermal states in the mechanical oscillator as shown in Fig.~\ref{fig:2}. We measure the qubit spectrum and compare the phonon distribution extracted from the deconvolution procedure to that expected for a thermal or displaced-thermal state. The good agreement between the phonon distributions as well as the associated qubit spectra substantiates the deconvolution procedure. 
In the dressed qubit spectra (Fig.~\ref{fig:2}a), individual sideband peaks cannot be resolved because these features are smeared by the large phonon number variance.
Nevertheless, features associated with those sidebands can be observed at the positions highlighted by the arrows.

\begin{figure} [h!]
    \centering
    \includegraphics{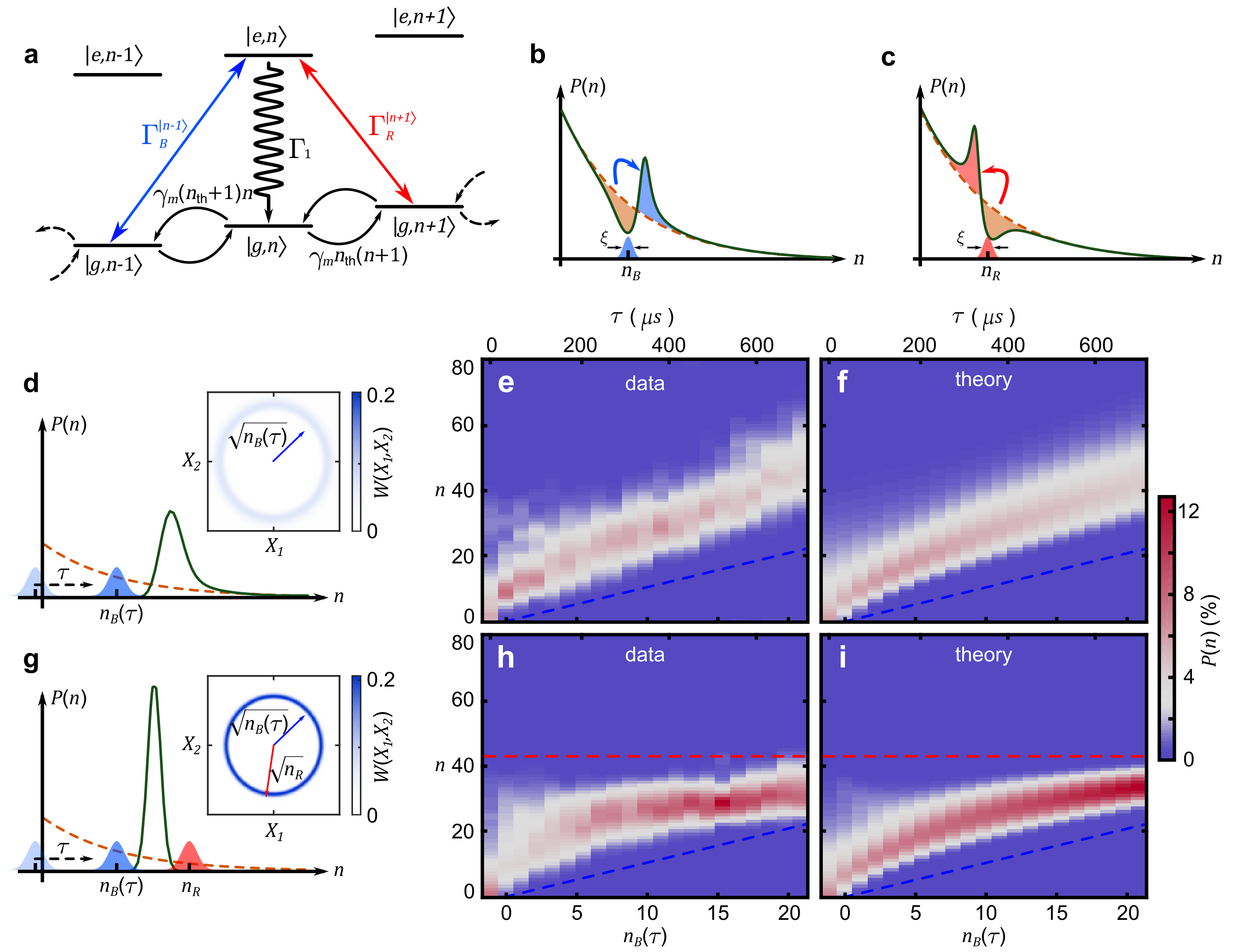}
    \caption{\textbf{Dissipative energy squeezing}}
    \label{fig:3}
\end{figure}

With the ability to extract the phonon distribution, we now use sideband transitions (Fig.~\ref{fig:3}a) to reduce the variance in phonon number. We squeeze the phonon population in Fock space (energy squeezing) by trapping it in between phonon-creating (blue) and phonon-annihilating (red) sideband transitions.
Continuously driving these sideband transitions alters the dissipative environment of the mechanical oscillator and changes the steady-state phonon distribution, as evident from the simplified\cite{Beaudoin2011DissipationUSC} system dynamics of Fig.~\ref{fig:3}a (also see supplementary material Sec.~VIIA). Because $\Gamma_2^*$ is much faster than all of the transition rates in these experiments, the final mechanical state contains no quantum coherence, and is fully described by the diagonal elements of its density matrix.
Similar to reference\cite{Viennot2018Phonon}, we adopt the technique\cite{Blais2007QuantumInformationProcessing} of driving ac-dither sidebands to access single-phonon sideband transitions (see supplementary material Sec.~VI). 
In contrast to that in conventional linear optomechanics\cite{Teufel2011GndStateCooling}, these transitions are crucially different in their phonon-number-sensitivity\cite{Viennot2018Phonon}.
Leveraging this feature, we address only a section of the phonon population with a characteristic width $\xi$ and displace it in number space (Fig.~\ref{fig:3}b-c). 
By slowly increasing the blue sideband drive frequency $\omega_B$ (chirping), we adiabatically move the center of the addressed transitions $n_{B}$ up in phonon space (Fig.~\ref{fig:3}d). In Fig.~\ref{fig:3}e, we show the effect of the chirp. Extracting the phonon population through reconstruction, we observe that the phonon population is emptied below $n_{B}$ and pushed to a higher occupation. 
In Fig.~\ref{fig:3}g-i, we squeeze the phonon population by turning on a red sideband drive centered on the transition $\ket{g,n_R} \leftrightarrow \ket{e,n_R-1}$, with $n_R$ close to but greater than the maximum value of $n_B$. A Fock state $\ket{n}$ will be cooled to a lower occupancy when the blue sideband transition rate is slower than the red sideband transition rate $\Gamma_B^{\ket{n}} < \Gamma_R^{\ket{n}}$, and vice versa. Thus, under conditions $\Gamma_B^{\ket{n_{B}}} > \Gamma_R^{\ket{n_{B}}}$ and $\Gamma_B^{\ket{n_R}} < \Gamma_R^{\ket{n_R}}$, a state that starts with $n_R > n > n_B$ cannot escape the bounds of the two sideband drives. Additionally, states with $n > n_R$ are eventually trapped between $n_R$ and $n_B$ by a combination of thermal equilibration and the action of the red sideband drive\cite{Viennot2018Phonon}.

\begin{figure} [h!]
    \centering
    \includegraphics{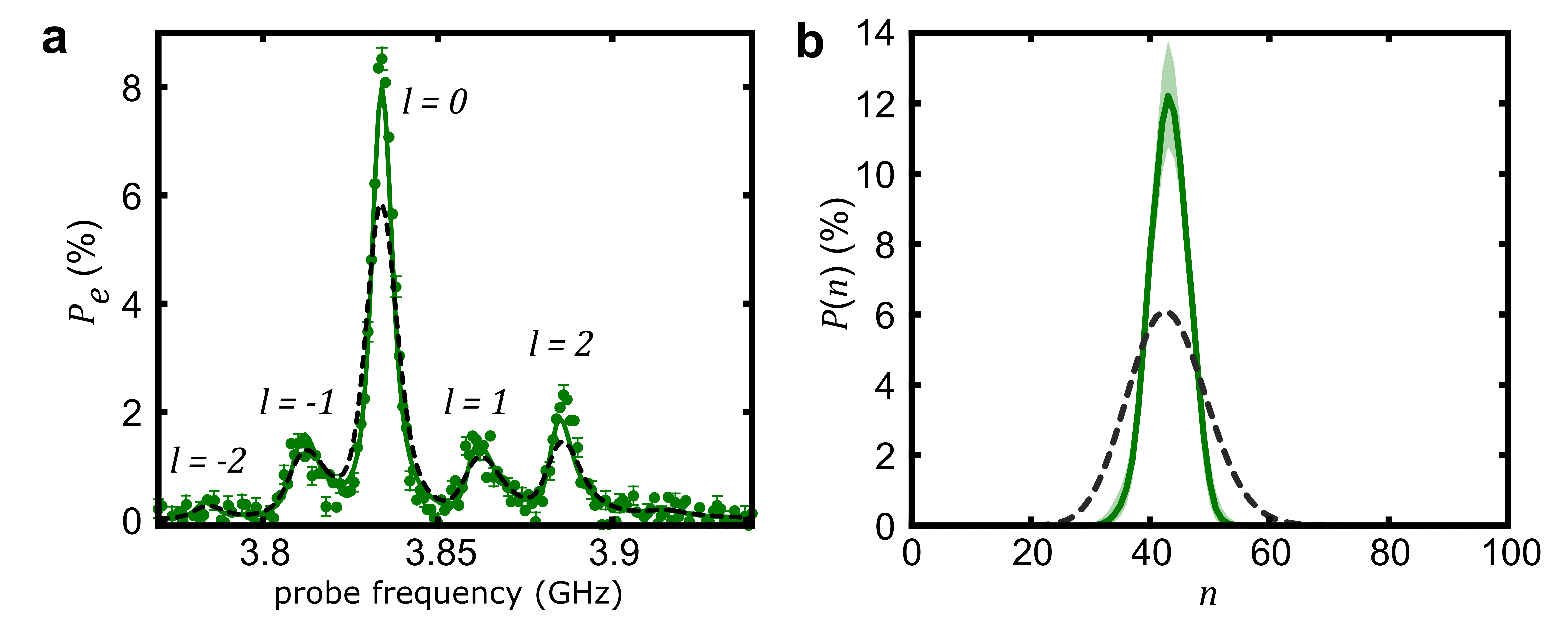}
    \caption{\textbf{Dissipatively stabilized sub-Poissonian state}}
    \label{fig:4}
\end{figure}

In Fig.~\ref{fig:4}, we use energy squeezing to prepare the mechanical oscillator in a nonclassical state. After optimizing the relative power and position of the two sideband drives, we squeeze the phonon population at mean phonon number $\braket{n} =43$ and prepare it in a sub-Poissonian state. We characterize the nonclassical nature of this state with the Fano factor $F = \text{var}(n)/\braket{n}$. For a Poisson distributed state, $F = 1$, and for a Fock state, $F = 0$. When $F < 1$, the phonon distribution is nonclassical\cite{Short1983ObserveSubPoissonian}, energy squeezed, and Fock-like. Extracting the phonon distribution through reconstruction, we find $F = 0.257_{-0.001}^{+0.002}$, where the bound is determined by the uncertainty in the bare qubit frequency. To quantify the confidence in the extracted Fano factor, we perform repeated reconstruction procedures on simulated experiments that have specified phonon distributions with $\langle n \rangle = 43$ and Fano factors $F_\text{true}$, as described in the supplementary material Sec.IX. For a given range of extracted Fano factor, $F_\text{extract}$, we can bound $F_\text{true}$. Specifically for $F_\text{extract}$ within the interval $[0.255,0.265]$, we find $F_\text{true} \leq 0.28$ with 95$\%$ confidence, and $F_\text{true} \leq 0.30$ with 99$\%$ confidence. 
This Fano factor can be related to a negativity in the Wigner function under the assumption of a Gaussian number distribution\cite{Lorch2014LaserRegime} (see supplementary material Sec.X).
We have thus demonstrated our ability to prepare a type of highly nonclassical mechanical state with large average energy but small fluctuations, quite distinct from quadrature squeezed states.
We choose to squeeze around $\braket{n} = 43$, where a spurious cooling effect (see supplementary material Sec.VIII) is small and the phonon population dynamics are more intuitive. However, because this method of energy squeezing creates states with a minimum width of $\xi \approx 7.1$ phonons independent of $\langle n\rangle$, it is conceivable to achieve a smaller $F$ by squeezing at higher $\braket{n}$.


In creating this energy-squeezed state, 
we can now resolve the sideband transitions that were obscured by the broad phonon distribution associated with the large thermal occupation in Fig.~\ref{fig:2}a. The center peak in the qubit spectroscopy ($l=0$ peak in Fig.~\ref{fig:4}a) corresponds to the qubit transition that conserves phonon number, whereas the satellite peaks at $\pm 2\omega_m$ ($l=\pm2$) are mostly associated with qubit transitions that create and annihilate pairs of phonons when the mechanical spring suddenly stiffens. 
Because charge noise creates a small random bias away from degeneracy, we also observe peaks at $\pm \omega_m$ associated with single-phonon sideband transitions (see supplementary material Sec.IIC). In contrast to the sidebands observed in reference\cite{Pirkkalainen2013HybridResonator} at $\langle n\rangle \sim 10^4$ phonons, these peaks are easily resolved at the relatively small phonon number $\langle n\rangle = 43$ because $\chi_m/\omega_m$ is much larger. Consequently, they can substantially alter the oscillator's phonon distribution.

The prominence of these sideband peaks also signifies the entry into a regime where the qubit's spontaneous decay can detectably alter the phonon distribution\cite{Beaudoin2011DissipationUSC} (see supplementary material Sec.VIIA).
In Fig.~\ref{fig:4}a, we measure the probability of exciting the qubit when driving at a particular frequency, but each peak must also correspond to a qubit decay process, driven by the quantum noise in the environment.
From the ratio of the area underneath the $l=0$ peak to the total area under all the peaks, we estimate that a qubit decaying from an initial state of $\ket{e,43}$ will only preserve the phonon number with 63\% probability! This probability will further diminish if $\chi_m$ were increased to better resolve individual phonon numbers.




In conclusion, we have demonstrated the preparation of nonclassical energy-squeezed states with quadratic electromechanics. We employed a dissipative stabilization technique that can simultaneously add energy to and extract entropy from a massive mechanical oscillator.
Requiring neither number-resolution nor coherent manipulation, this technique provides a hardware-efficient and accessible path toward creating highly energized Fock-like states in other cQED experiments. Such states are
resources for quantum metrology. They have been analyzed for their ability to improve the sensitivity of gravitational wave detectors\cite{Caves1980WeakClassicalForceMeasurement,Braginsky1980QNDelectromechanics} and demonstrated to resolve small forces on trapped ions\cite{Wolf2019FockStateForceDetection}.
Indeed, the sub-Poissonian state we create reveals that a spontaneous qubit transition is likely to alter the occupation of the harmonic oscillator, a striking feature in the regime of $n\chi_m \sim \omega_m$. To reach this regime, we engineered an ultra-strong\cite{FriskKockum2019UltrastrongMatter} electromechanical coupling rate of $g_m \approx 2\pi \times 22$~MHz, limited by our ability to readout the qubit state\cite{Viennot2018Phonon}. Surpassing this limitation would allow us to reach the phonon-number-resolving regime for arbitrary quantum control over mechanical oscillators, and to observe the ultra-strong coupling induced virtual phonons in the mechanical ground state\cite{Ashhab2010Qubit-oscillatorStatesUSC,FriskKockum2019UltrastrongMatter}.  


\bibliographystyle{naturemag} 
\bibliography{Bibliography.bib}

\subsection*{Data and code availability}
The data and codes that support the plots within this paper and other findings of this study are available from the corresponding author upon reasonable request.

\acknowledgements{We thank Daniel Palken, Lucas Sletten, Robert Delaney, Felix Beaudoin and Lin Tian for fruitful discussions. We gratefully acknowledge Ray Simmonds and Florent Lecocq for their help with the fabrication of the device. We thank Maxime Malnou and Daniel Palken for providing us with a Josephson parametric amplifier. We acknowledge funding from National Science Foundation (NSF) under Grant No. 1734006. J. J. Viennot acknowledges financial support from the European Union's H2020 program under the Marie Sklodowska-Curie Grant No. 841618}

\subsection*{Competing Financial Interests statement}
The authors declare no competing financial interests. 

\newpage
\nolinenumbers

\noindent \textbf{a,} The capacitances of two mechanically compliant capacitors, $C_m^{\pm}(x)$, are modulated with opposite phase by the anti-symmetric motion of the mechanical oscillator (blue).
An applied voltage $V_\text{dc}$ converts this modulation of capacitance to a voltage $V_g(x)$
across the open terminals A and B. \textbf{b,} The Thevenin equivalent representation of the circuit seen by the junctions is a Cooper-pair box qubit, with a mechanical-position dependent gate charge $n_g(x) = V_g(x) \times C_g/2e$ that \textbf{c,} tunes the qubit energy $E_\text{CPB}$. 
The qubit ground ($\ket{g}$) and excited ($\ket{e}$) states are superpositions of two charge states (dashed lines) of the circuit differing by one Cooper-pair, with the average value of the ground and excited state energy defined to be 0. These energies are linearly dependent on $x$ with slope $\pm 2\hbar g_m/x_{zp}$, defining $g_m$ as the qubit-mechanics coupling rate. The degeneracy between the charge states at $n_g = 1/2$ (charge degeneracy point) is lifted by the tunneling of Cooper pairs across the junctions at rate $E_J/\hbar$. At charge degeneracy, the qubit transition frequency senses the square of mechanical displacement with quadratic coupling strength $(2g_m)^2/(E_J/\hbar)$. \textbf{d,} False-colored scanning electron micrograph (at an angle) of the micromechanical oscillator (blue) suspended above two electrodes (green and yellow) to form mechanically compliant capacitors. The dc bias line imposes a voltage onto the oscillator plate. \textbf{e,} Top view of the device. The two bottom electrodes are shunted by two Josephson junctions (JJ) in parallel to form a superconducting qubit. \textbf{f,} A qubit excitation causes a sudden change of the mechanical potential (parabolas) and a non-zero overlap between spatial wavefunctions (lines) of different mechanical states of opposite qubit excitation. Because of symmetry, this process only connects an initial state $\ket{g,n}$ (shaded green) with states of the same mechanical parity $\ket{e,n\pm2l}$ (shaded red), creating or annihilating phonons by pairs. 

\newpage

\noindent 
\textbf{a}, The qubit spectrum (dots) is measured when the mechanical oscillator is in a thermal state (orange) or a displaced thermal state (purple). Although individual sideband peaks are not resolvable because of the broad phonon distribution, we can still observe features associated with them at the positions highlighted by the vertical arrows. 
We perform least-squares fit on the thermal state qubit spectrum (orange) using Eqn.(\ref{eq:DressedSpectrum}) assuming a thermal distribution (orange dashed) with only two free parameters: $n_\text{th} = 17.7$ and the bare qubit frequency (see supplementary material Sec.~IIID). Holding these parameters fixed and assuming a displaced thermal state distribution, we also fit the purple data to find the only free parameter  $n_\text{disp} = 43.3$.
\textbf{b}, Alternatively, we can extract the phonon distributions without assuming a particular form using a deconvolution procedure (solid lines) and find their 90\% confidence intervals (shaded areas) using non-parametric bootstrapping. For comparison, we also plot the associated qubit spectra (solid lines) in \textbf{a}, and the fitted phonon distributions (dashed lines) in \textbf{b}. The presented data is representative of other displaced thermal states we measure, where we confirm the extracted $n_\text{disp}$ scales linearly with the power of the displacement drive (see supplementary material Sec.~V).

\newpage
\noindent  \textbf{a,} When the blue sideband transition $\ket{g,n-1} \leftrightarrow \ket{n,e}$ is driven continuously at rate $\Gamma_{B}^{\ket{n-1}}$ and the qubit decay rate is the dominant relaxation process ($\gamma_m \ll \Gamma_1$), the combination of the two results in the addition of one phonon in the mechanical oscillator at rate $\Gamma_B^{\ket{n+1}}$. ($\Gamma_1 > \Gamma_B^{\ket{n-1}}$ for this work). 
Similarly, a red sideband drive removes one phonon at rate $\Gamma_R^{\ket{n+1}}$.
\textbf{b, c,} The sideband drives are phonon-number-sensitive, addressing a section of the phonon population centered at $n_B$ or $n_R$ (for blue or red sideband drives) with a characteristic width of $\xi \approx 7.1$ phonons. 
A blue (\textbf{b}) or red (\textbf{c}) sideband drive applied on an initial thermal state (dashed orange) creates a distortion in the phonon population (solid green) at time scale $\big( n_B/ \Gamma_B^{\ket{n_B}} \big)$ or $\big( n_R/\Gamma_R^{\ket{n_R}} \big)$ .
\textbf{d,} Alternatively, adiabatically increasing the blue sideband drive frequency (chirping) should empty all phonon population below the final value of $n_B$. The inset shows that the resulting state's Wigner function $W(X_1,X_2)$ is a narrow ring around the ($X_1$,$X_2$) quadrature space origin with inner-radius approximately given by $\sqrt{n_B}$. 
\textbf{e,} Chirping the sideband drive to a final position of $n_B(\tau)$ (bottom-axis) by stopping the chirp at time $\tau$ (top-axis), we measure the qubit spectrum (see supplementary material Fig.11S) and extract the phonon distribution (color-scale vs. y-axis) using deconvolution. At any time, the population is empty below the line $n = n_B(\tau)$ (dashed blue). 
\textbf{f,} Using a master equation calculation, we find the expected phonon distribution for the chirping protocol used in \textbf{e} (see supplementary material Sec.~VII).
\textbf{g,} When the blue sideband drive is chirped toward a static red sideband drive centered at $n_R$, the phonon population should be trapped in between, and squeezed in number space. Inset: The Wigner function of the resulting energy squeezed state is non-Gaussian and radially symmetric about the quadrature space origin, quite distinct from a quadrature squeezed state.
\textbf{h, i,} As in \textbf{e} and \textbf{f}, We compare the measured (\textbf{h}) and the expected (\textbf{i}) phonon distribution for this energy squeezing protocol with $n_R = 44$ (red dashed).

\newpage
\noindent The qubit spectrum (\textbf{a}, dots) is measured after an optimized energy squeezing protocol. The standard error of the mean (SEM) are extracted individually for each frequency point (see supplementary material Sec. IX), and plotted as error-bars for every fourth point.
The phonon distribution (\textbf{b}, solid green) with its 90\% confidence interval (shaded green) is extracted from a deconvolution procedure, and is found to be sub-Poissonian and characterized by $F = \text{var}(n)/\braket{n} = 0.257 <1$. For reference, the phonon distribution and the expected qubit spectrum are plotted for a coherent state (Poissonian, dashed black). In the qubit spectrum (\textbf{a}), the sub-Poissonian nature of the mechanical state is evident in the narrower lineshape. The peaks visible are separated by $\omega_m$, and corresponds to transitions $\ket{g,n} \rightarrow \ket{e,n + l}$.


\end{document}


\title{Supplementary Information}
\maketitle

\hypersetup{linkcolor=black}
\tableofcontents
\hypersetup{linkcolor=red}

\section{Device and equivalent circuit} \label{sec:System}

\subsection{Device overview} \label{sec:Full}

\begin{figure} [h!]
    \centering
    \includegraphics{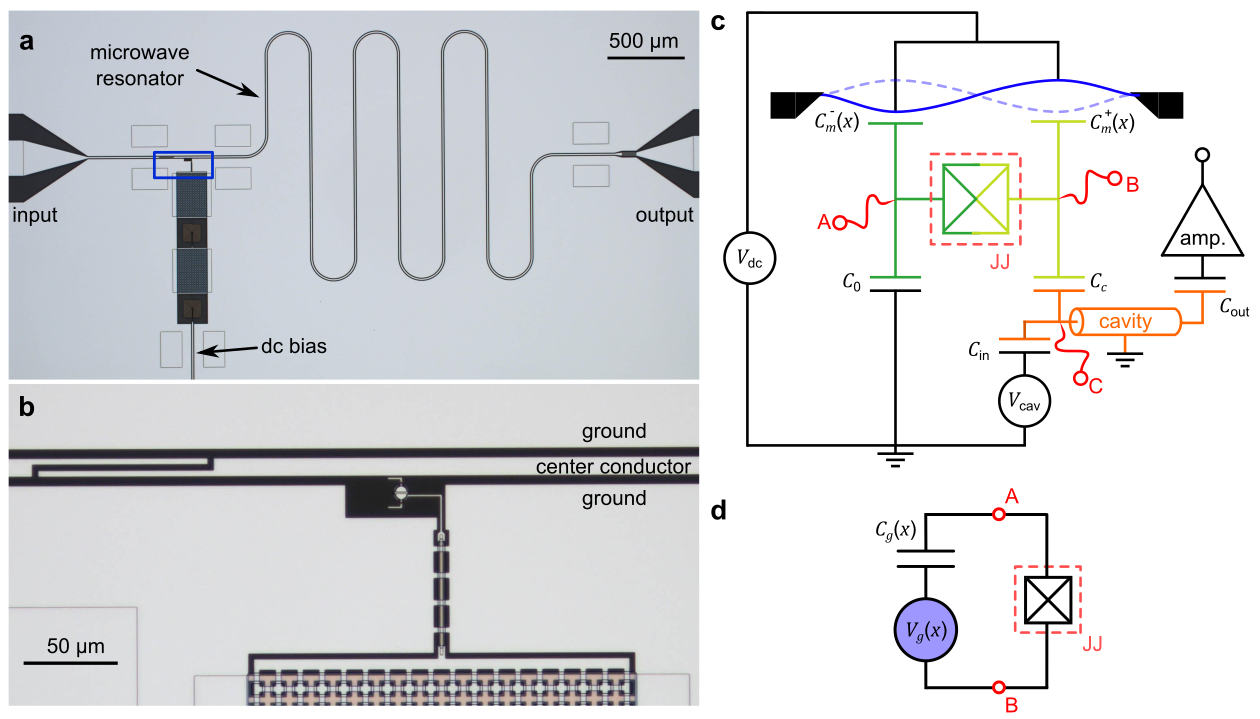}
    \caption{\textbf{The device}\\
    \textbf{a,} Optical micrograph of the device chip shows the CPW resonator (microwave resonator), input and output ports and the dc bias port. A on-chip LC filter, composed of inductance in series with waffled capacitance to ground, can be seen on the dc bias line.
    \textbf{b,} A zoom on the blue rectangle of \text{a} shows the qubit-mechanics device embedded in between the center conductor and one ground plane of the CPW resonator. The input coupling capacitance $C_\text{in}$ of the resonator is visible on the left. 
    \textbf{c,} 
    A voltage $V_\text{dc}$ is applied on the suspended Aluminum disk (blue). Underneath the disk, two bottom eletrodes placed at the motional anti-node of the disk form two mechanically compliant capacitors $C_m^\pm (x)$. The two superconducting islands (green and yellow) are connected by Josephson junctions. One of the two islands is capacitively coupled to a microwave resonator (orange) that allows for readout and control of the qubit. To read out the qubit, we measure the microwave resonator in transmission, and the output port of the cavity is connected to the amplifier chain through coupling capacitor $C_\text{out}$. For low frequency signals around $\omega_m$, node C can be treated as if it were grounded, and we recover Fig.1(c) of the main text. The anti-symmetric motion converts $V_\text{dc}$ into a voltage $V_g(x)$ across the open terminals A and B.     
    \textbf{d,} The Thevenin equivalent circuit around the Josephson junctions is a Cooper-pair box qubit. This qubit is coupled to mechanical motion through a position-dependent gate-charge $n_g(x) = C_g(x)V_g(x)/2e$.}
    \label{fig:device micrograph}
\end{figure}

Photograph of the full device is given in Fig.\ref{fig:device micrograph}(a). A dc bias port is used to apply an external dc voltage $V_\text{dc}$ on the suspended Aluminum disk that causes the qubit-mechanics coupling. It also allows rf frequency drives up to 260~MHz, and is used extensively for classical ac modulation of the gate-charge (Sec.\ref{sec:qubit phonon responce}), coherent driving of the mechanical oscillator (Sec.\ref{sec:coherent drive}), and providing the ac-dither for the sideband drives (Sec.\ref{sec:acDither}). To prevent the qubit energy from decaying through the dc bias line, a on-chip low-pass filter with a cut-off frequency of 1~GHz is incorporated. A co-planar waveguide (CPW) resonator provides the ability to control and readout the qubit. As in many standard circuit Quantum Electrodynamics (cQED) setups\cite{Wallraff2004StrongElectrodynamics.}, we dispersively couple the qubit to the CPW resonator by inserting it between the center conductor and one ground plane, as shown in Fig.1S(b). In the following, the microwave resonator is also referred to as the cavity, to avoid possible confusion with the mechanical oscillator.

The device can be related to the circuit schematic of \textbf{Fig.}\ref{fig:device micrograph}(c). A voltage $V_\text{dc}$ is imposed on the suspended disk. Two bottom electrodes, placed at the motional anti-nodes of the disk, are connected by Josephson junctions. One of the electrodes is connected to the ground plan through capacitance $C_0$. The other electrode is capacitively coupled to the microwave resonator through capacitance $C_c$, which allows for coherent control and readout of the qubit.

\begin{figure} [h!]
    \centering
    \includegraphics{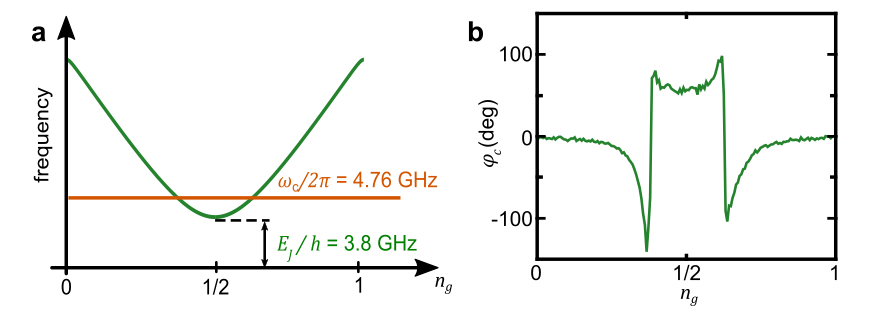}
    \caption{\textbf{Cooper-pair box qubit}\\
    \textbf{a,} The qubit frequency (green) is strongly dependent on the gate-charge $n_g$, and would cross the cavity at $n_g = 0.38$ and $0.62$ if they were uncoupled.
    \textbf{b,} Because they are coupled, the qubit state alters the cavity frequency. 
    The phase $\varphi_c$ of the cavity transmission plotted versus $n_g$ shows that the qubit ground state shifts the cavity frequency. 
    At a fixed value of $n_g$, this phase enables qubit readout.}
    \label{fig:CPB}
\end{figure}

In this experiment, we use a charge qubit in the Cooper-pair box (CPB) regime\cite{Bouchiat1998QuantumPair,Makhlin2001Quantum-stateDevices}. Compared to a transmon, the highly anharmonic energy levels of a CPB provides a much stronger sensitivity to low frequency charge fluctuations\cite{Koch2007Charge-insensitiveBox,Gely2018NatureFluctuations}. The Hamiltonian of the qubit is given by
\begin{equation} \label{eqn:CPB}
    H_q = 4E_c(\hat{n}-n_g)^2-E_J \cos{(\hat{\phi})},
\end{equation}
where $E_c$ and $E_J$ are the charging and Josephson energy respectively ($E_c \approx E_J$), $\hat{n}$ is the Cooper-pair number operator and $\hat{\phi}$ is the superconducting phase operator. We take the lowest two energy levels to function as the qubit, and its frequency is strongly dependent on gate-charge $n_g$,
\begin{equation}\label{eqn:qubit frequency}
    \omega_q = \sqrt{E_J^2+(4E_c)^2(1-2n_g)^2},
\end{equation}
as shown in \textbf{Fig.}\ref{fig:CPB}(a). Because of this strong dependence, noise in gate-charge can significantly change $\omega_q$ and quickly decohere the qubit. To alleviate this problem, we need to operate the qubit at the charge degeneracy point $n_g = 1/2$, where the qubit frequency is to first order insensitive to $n_g$. We achieve this by applying a dc voltage on port $V_\text{cav}$.

To readout the qubit state, we dispersively couple the microwave resonator (i.e., cavity) to the qubit and measure its phase in transmission. 
When the cavity frequency is pulled by the state of the qubit\cite{Blais2004CavityComputation}, it causes the cavity transmission phase $\varphi_c$ to change. \textbf{Fig.}\ref{fig:CPB}(b) shows $\varphi_c$ measured at the uncoupled cavity frequency as a function of $n_g$ when the qubit is in the ground state, $\sigma_z = -1$. When the qubit is excited, the $\varphi_c$ has the opposite sign. Because of the short qubit lifetime $T_1$, the qubit readout is not "single-shot". Instead, we adopt an incoherent measurement technique, and infer the average occupation of the qubit $\langle \hat{\sigma}_z \rangle$ through the cavity transmission phase\cite{Schuster2005AcField}. Operating around $n_g = 1/2$, the qubit excited state probability is,
\begin{equation}\label{eqn:Pe}
P_e  = \frac{1}{2}(1+\sigma_z) \approx \frac{1}{2} - \varphi_c/2 \varphi_{n_g=0.5}^{\ket{g}}.
\end{equation}

\begin{figure} [h!]
    \centering
    \includegraphics{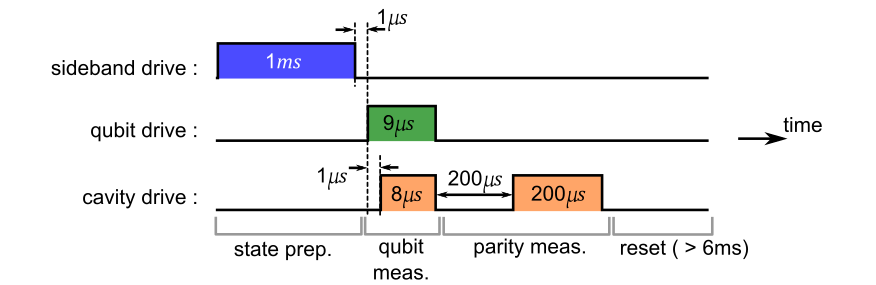}
    \caption{\textbf{A sample pulse sequence}\\
    We sketch the pulse sequence (not to scale) used to measure Fig.4 of main text. We start by preparing the desired mechanical state with sideband drives (blue). After waiting for the qubit to decay back to the ground state, we measure the qubit spectrum by driving the qubit (green) and measuring the cavity (orange) phase $\varphi_c$. To ensure we measure the steady-state qubit spectrum, the qubit drive is applied $1~\mu$s before the start of the phase measurement.
    After waiting for $200~\mu$s, we again measure $\varphi_c$ when the qubit is in the ground state. Post-selecting on this phase, we ensure the qubit remains close to the charge-degeneracy point despite possible quasi-particle tunneling events (i.e. correct qubit parity). 
    Finally, we wait for more than $6~$ms for the mechanical oscillator to thermalize back to its environment before starting the next measurement cycle.}
    \label{fig:Protocal}
\end{figure}

As an example, here we describe the pulse sequence (Fig.~\ref{fig:Protocal}) that resulted in Fig.4 of main text: we start by preparing the mechanical oscillator in the desired state by pulsing on simultaneously the sideband drives and the ac-dither for $1~$ms. Because $T_1 \approx 0.26~\mu$s, we wait $1~\mu$s after the sideband pulse for the qubit to decay back to the ground state. We then measure the qubit spectrum by driving the qubit at various frequencies and measuring the cavity transmission phase for a short $8~\mu$s to avoid measurement backaction (see Sec.\ref{sec:backaction}). To ensure we measure the steady-state qubit spectrum, the qubit drive is applied $1~\mu$s before the start of the phase measurement. We finally measure the qubit parity\cite{Viennot2018Phonon}, and wait more than $6~$ms for the mechanical oscillator to thermalize back to its environment before starting the next measurement cycle.

\begin{table}[h!]
\begin{center}
 \begin{tabular}{m{8cm} m{8cm}} 
 \hline \hline
 Parameter & Value\\
 \hline
qubit Josephson energy & $E_J/\hbar \approx 2\pi \times 3.8$~GHz\\
qubit charging energy & $E_c/\hbar \approx 2\pi \times 2.9$~GHz\\
qubit lifetime & $T_1 \approx 0.26 \mu$s\\
qubit intrinsic decoherence time & $T_2^* \approx 0.08  \mu$s\\
mechanical frequency & $\omega_m = 2\pi \times 25$~MHz\\
mechanical thermal occupation & $n_{th} = 13 \sim 20$\\
mechanical damping rate & $\gamma_m \approx 2\pi \times 94$~Hz\\
mechanical thermal decoherence rate & $n_{th}\gamma_m \approx 2\pi \times (1.2 \sim 1.9)$~kHz\\
cavity (microwave resonator) frequency &$\omega_c \approx 2\pi \times 4.76$~GHz\\
cavity (microwave resonator) linewidth & $\kappa \approx 2\pi \times 2.3$~MHz\\
single phonon ac Stark shift & $2\chi_m \approx 2\pi \times 0.52$~MHz ($V_\text{dc} = 6$~V, $n_g = 1/2$)\\
qubit-mechanics coupling rate & $g_m \approx 2\pi \times 22$~MHz ($V_\text{dc} = 6$~V, $n_g = 1/2$)\\
qubit-cavity coupling rate & $g_c \approx 2\pi \times 37$~MHz ($n_g = 1/2$)\\
 \hline
\end{tabular}
\end{center}
\caption{Essential parameters of the system} \label{table:2S} 
\end{table}

\subsection{Thevenin equivalent circuit and qubit-mechanics interaction}\label{sec:EquivalentCircuit}

In this section we derive the coupling between the qubit and the mechanical oscillator which arises from a motional modulation of the CPB gate-charge. To see this, we analyze the circuit in in \textbf{Fig.}\ref{fig:device micrograph}(c) for low frequency signals that are close to the mechanical resonance or at dc. At such frequencies, because the linear capacitance of the microwave resonator $C_\text{shunt}$ is much larger than the other capacitances ($C_\text{shunt} \gg C_\text{in}, C_\text{out}, C_c$), node C in \textbf{Fig.}\ref{fig:device micrograph}(c) can be treated as if it were grounded. Thus, we recover Fig.1(c) of the main text.
Seen by the junctions, the Thevenin equivalent representation of the circuit is shown in \textbf{Fig.}\ref{fig:device micrograph}(d). The equivalent voltage $V_g(x)$ is the voltage difference across open terminals A and B,
\begin{equation}
\label{eqn:Veq_full}
\begin{aligned}
    V_g(x) = V_A(x)-V_B(x) &= V_\text{dc}\left(\frac{C_m^-(x)}{C_m^-(x)+C_0}-\frac{C_m^+(x)}{C_m^+(x)+C_c}\right),
\end{aligned}
\end{equation}
with $C_m^\pm = C_m^0/(1\pm \frac{x}{x_0})$, where $x_0$ is the static separation between the suspended disk and the bottom electrodes.
Although we intentionally create an asymmetry between $C_c$ and $C_0$ to allow for coherent driving of the mechanical oscillator (see Sec.\ref{sec:coherent drive}), this asymmetry only introduces minor corrections to the coupling rate but doesn't change the form of the coupling.
Thus for clarity, we treat the symmetric case here with $C_0 = C_c$.
Expanding around small motion at $x=0$, eqn.(\ref{eqn:Veq_full}) simplifies to its approximate form,
\begin{equation}\label{eqn:Veq}
    V_g(x) \approx 2\frac{ C_m^0 C_0}{(C_m^0+C_0)^2}\frac{x}{x_0}V_\text{dc} + \mathcal{O}\left(\frac{x}{x_0}\right)^2.
\end{equation}
Similarly, the equivalent capacitance $C_g(x)$ is found by replacing the dc voltage source with a short circuit,
\begin{equation}
    C_g(x) \approx \frac{1}{2}(C_0+C_m^0) +  \mathcal{O}\left(\frac{x}{x_0}\right)^2.
\end{equation}
Therefore, to first order in $x$, the circuit reduces to a Cooper-Pair box qubit, whose gate-voltage is linearly controlled by the mechanical displacement, as we have illustrated in Fig.1(d) of the main text.

The position-dependent gate-voltage and capacitance of the CPB leads to a qubit-mechanics coupling through the gate-charge, $n_g(x) = C_g(x) \times V_g(x)/2e$. For small motional amplitudes that we are concerned with in this experiment (tens to hundreds of motional quanta), the modulation in gate-charge is always much smaller than a single Cooper-pair. Therefore, we can confine the CPB charge basis to two adjacent charge states. The Cooper-pair number operator in eqn.(\ref{eqn:CPB}) reduces to $\hat{n} = (\hat{\sigma}_z^\prime + 1)/2$, where $\hat{\sigma}_z^\prime$ points along the $E_c$ axis. Defining $\hat{\sigma}_z$ to align with the energy quantization axis,
the interaction Hamiltonian is given by,
\begin{equation}\label{eqn:coupling}
    H_I = \hat{x} \left. \frac{\partial}{\partial x}H_q \right\vert_{n_g=1/2} = 4E_c\frac{\partial n_g(x)}{\partial x}x_{zp}(\hat{a}+\hat{a}^\dagger)\big(\cos{\theta_0} \hat{\sigma}_x + \sin{\theta_0} \hat{\sigma}_z\big) = \hbar g_m \big (\hat{a}+\hat{a}^\dagger)(\cos{\theta_0} \hat{\sigma}_x + \sin{\theta_0} \hat{\sigma}_z\big),
\end{equation}
where $\hat{x} = x_{zp}(\hat{a}+\hat{a}^\dagger)$ is the mechanical position operator, $x_{zp} = \sqrt{\frac{\hbar}{2m\omega_m}}$ is the mechanical zero point motion, and $\theta_0 = \text{arctan}[4E_c(1-2n_g)/E_J]$ is the mixing angle between charging energy and Josephson energy. The single phonon qubit-mechanics coupling rate is given by
\begin{equation}\label{eqn:coupling rate}
    g_m = \frac{4E_c}{\hbar}\frac{\partial n_g(x)}{\partial x}x_{zp} = \frac{4E_c}{2e\hbar} \frac{C_m^0 C_0}{C_m^0 + C_0}\frac{x_{zp}}{x_0}V_\text{dc}.
\end{equation}

\section{Effective Interaction}\label{sec:Interaction}
\subsection{Dispersive transformation and effective quadratic coupling}\label{sec:QuadraticCoupling}

The qubit-mechanics interaction in eqn.(\ref{eqn:coupling}) becomes an effective quadratic interaction under dispersive transformation. Operating at the charge degeneracy point $n_g=1/2$ and taking $\hbar=1$, the Hamiltonian of the qubit-mechanics system is given by,
\begin{equation}\label{eqn:bare Hamiltonian}
    H_0 = \omega_m \hat{a}^\dagger \hat{a} + \frac{1}{2}\omega_q \hat{\sigma}_z + g_m\hat{\sigma}_x(\hat{a}+\hat{a}^\dagger).
\end{equation}
Because of both the ultra-strong coupling strength ($g_m / \omega_m \approx 0.9$), and the large difference in the resonant frequencies ($\omega_m / \omega_q \approx 0.006$), the counter rotating terms $\sigma_+a^\dagger$ and $\sigma_- a$ contribute significantly. Therefore, we cannot apply the Rotating Wave Approximation (RWA) to simplify eqn.(\ref{eqn:bare Hamiltonian}) into the typical Jaynes-Cummings Hamiltonian. Nevertheless, the large difference between the coupling rate and the qubit frequency ($\omega_q \gg g_m$) means that there is no spontaneous transfer between qubit and mechanical excitation\cite{Larson2007DynamicsBottles}. Following reference\cite{Zueco2009Qubit-oscillatorApproximation}, applying a unitary dispersive transformation,
\begin{equation}\label{eqn:dispersive transformation}
    \hat{U}_\text{disp} = \text{exp}\left[\frac{g_m}{\Delta}(\hat{a} \hat{\sigma}_+ - \hat{a}^\dagger \hat{\sigma}_-) + \frac{g_m}{\Sigma}(\hat{a}^\dagger \hat{\sigma}_+ - \hat{a} \hat{\sigma}_-) \right],
\end{equation}
where $\Delta = \omega_q - \omega_m$ and $\Sigma = \omega_q + \omega_m$, and keeping terms to the first order of $g_m/\Delta$ and $g_m/\Sigma$, we find
\begin{equation}\label{eqn:dispersive Hamiltonian}
    H_\text{disp} = \omega_m \hat{a}^\dagger \hat{a} + \frac{1}{2}\omega_q \hat{\sigma}_z + \frac{1}{2}\chi_m \hat{\sigma}_z (\hat{a}+\hat{a}^\dagger)^2,
\end{equation}
where $\chi_m = g_m^2 \big( 1/\Delta + 1/\Sigma \big)$ includes the Bloch-Siegert shift\cite{Bloch1940MagneticFields,Zueco2009Qubit-oscillatorApproximation,Beaudoin2011DissipationQED}. 

Because a qubit excitation is much faster than the mechanical dynamics, the position can be regarded as stationary under a sudden qubit excitation. The physics is clear if the Hamiltonian is written as
\begin{equation}\label{quadratic coupling}
    H_{\text{eff}} = \frac{1}{2}\omega_q \hat{\sigma}_z + \frac{\hat{p}^2}{2m}  + \frac{1}{2}k\hat{x}^2 + \frac{1}{2} k \bigg(\frac{2 \chi_m }{\omega_m} \bigg) \hat{\sigma}_z \hat{x}^2,
\end{equation}
where we recover the effective quadratic coupling Hamiltonian in the main text by replacing the phonon number operators with the position and momentum operators. Here,
$\hat{x} = \sqrt{\frac{1}{2m \omega_m}}(\hat{a}^\dagger+\hat{a})$, and $\hat{p} = i \sqrt{\frac{m \omega_m}{2}}(\hat{a}^\dagger-\hat{a})$ are position and momentum operators respectively. With the Hamiltonian in this form, the qubit-mechanics coupling can be expressed as a qubit-state dependent spring constant,
\begin{equation}
    k(\hat{\sigma}_z) = k\bigg(1+ \frac{2\chi_m}{\omega_m} \hat{\sigma}_z\bigg).
\end{equation}

\subsection{System eigenstates and qubit excitation}\label{sec:Eigenstates}

In this section, we describe theoretically the effect of a qubit state dependent spring constant. We show that the qubit spectrum acquires sideband transitions that alter the phonon number and a dispersive shift proportional to the phonon number. We use both a Frank-Condon description (Sec.\ref{sec:FrankCondon}) and direct diagonalization (Sec.\ref{sec:USC}) to derive these effects. In particular, we diagonalize the system Hamiltonian in the presence of a coherent qubit drive to predict the qubit spectrum. 

\subsubsection{Frank-Condon description}\label{sec:FrankCondon}

The impedance $Z_m = \sqrt{k m}$ determines the spatial scale of the mechanical wavefunction. For a given qubit state $\sigma_z$, the mechanical spatial wavefunction for a phonon Fock state $\ket{n}$ is given by\cite{Shankar},
\begin{equation}\label{eqn:wavefunction}
    \psi(x,n,\sigma_z) = \frac{1}{\sqrt{2^n n!}}\left(\frac{Z_m^{\sigma_z}}{\pi\hbar}\right)^\frac{1}{4}e^{Z_m^{\sigma_z} x^2/2\hbar}H_n\left(x \sqrt{\frac{Z_m^{\sigma_z}}{\hbar}}\right),
\end{equation} 
where $Z_m^{\sigma_z} = \sqrt{k(\sigma_z)m}$ is the qubit-state dependent mechanical impedance, and the functions $H_n(z)$ are Hermite polynomials.

As such, a qubit excitation can connect otherwise orthogonal mechanical states, according to the Frank-Condon principle\cite{Franck1925ELEMENTARYREACTIONS,Condon1926ASYSTEMS}. The probability of such a transition is given by the overlap in the spatial wavefunctions,
\begin{equation}\label{eqn:P}
    P_{n,m} \propto \int_{-\infty}^{\infty} \psi^*(x,n,-1)\psi(x,m,1) dx,
\end{equation}
where $P_{n,m}$ is the probability of observing transition $\ket{g,n}\rightarrow \ket{e,m}$. Because of the symmetry in the mechanical potential, we expect only transitions that change the phonon occupation by an even number. 
When $m=n$, this qubit transition preserves the mechanical phonon number. However, when $m>n$, we have a blue sideband transition, where phonons are added while exciting the qubit. Vice versa, when $m<n$, a red sideband transition is realized, which cools the mechanical oscillator while exciting the qubit. 

\subsubsection{Direct diagonalization}\label{sec:USC}
To directly diagonalize the Hamiltonian of eqn.(\ref{eqn:dispersive Hamiltonian}), we introduce unitary transformation\cite{Joshi2017Qubit-flip-inducedModel,Machado2019QuantumOptomechanics},

\begin{equation}\label{eqn:squeezing}
    \hat{S}\bigg(\mathcal{r}(\hat{\sigma}_z)\bigg) = \text{exp}\left[\frac{1}{2}\mathcal{r}(\hat{\sigma}_z)(\hat{a}^2- \hat{a}^{\dagger^2}) \right],   
\end{equation}
where, 
\begin{equation}\label{eqn: squeezing amp}
    \mathcal{r}(\hat{\sigma}_z) = \frac{1}{2}\text{arctanh} \left( \frac{\chi_m \hat{\sigma}_z}{\omega_m + \chi_m \hat{\sigma}_z} \right) \approx \frac{\chi_m}{2\omega_m}\hat{\sigma}_z \equiv r\hat{\sigma}_z.
\end{equation}
This unitary transformation diagonalizes the Hamiltonian into the familiar form of a dispersive Jaynes-Cummings Hamiltonian, 
\begin{equation}\label{eqn: diagnoal Hamiltonian}
    H_{sq} = \sum_n \bigg\{ \omega_n^- \ket{g}_b   \ket{n}_b \prescript{}{b}{\bra{n}}   \prescript{}{b}{\bra{g}}
    + \omega_n^+ \ket{e}_b   \ket{n}_b \prescript{}{b}{\bra{n}}    \prescript{}{b}{\bra{e}} \bigg\}
\end{equation}
where $\ket{g}_b$, $\ket{e}_b$, and $\ket{n}_b$ are eigenstates of the the uncoupled qubit and mechanical oscillator respectively, and the eigenvalues are given by $\omega_{n}^{\pm} = n\omega_m \pm \frac{1}{2}(\omega_q+2\chi_m n) = n\omega_m \pm \frac{1}{2}\omega_q^n$, with
\begin{equation}\label{eqn:stark shift quantum}
    \omega_q^n = (\omega_q^b + \chi_m) + 2\chi_m n
\end{equation}
the phonon number dependent qubit frequency, and $\omega_q^b$ the bare qubit frequency. The eigenstates of the diagonalized Hamiltonian $\ket{g}_b   \ket{n}_b$ and $\ket{e}_b   \ket{n}_b$ are related to the the eigenstates of the original Rabi Hamiltonian eqn.(\ref{eqn:bare Hamiltonian}) through the unitary transformation $\hat{S}\bigg(\mathcal{r}(\hat{\sigma}_z)\bigg)$,
\begin{equation}\label{eqn:eigenstate main text}
\begin{aligned}
    \ket{g,n} &= \hat{S} (r\hat{\sigma}_z) \ket{g}_b   \ket{n}_b = \ket{g}_b  \hat{S}(-r)\ket{n}_b,\\
    \ket{e,n} &= \hat{S} (r\hat{\sigma}_z) \ket{e}_b   \ket{n}_b = \ket{e}_b  \hat{S}(r)\ket{n}_b.
\end{aligned}
\end{equation}
Indeed compared to the uncoupled mechanical oscillator, the excited state qubit stiffens the mechanical spring, and squeezes the oscillator energy along its position axis. Similarly, the qubit ground state loosens the mechanical spring, and anti-squeezes the oscillator energy along its position axis. 

With the Hamiltonian diagonalized, we can now understand how a coherent drive on the qubit will affect the system. Already from eqn.(\ref{eqn:eigenstate main text}), we can observe that under a qubit excitation, the mechanical occupation will not be conserved,
\begin{equation}\label{eqn:mech state overlap}
    \alpha_{mn} = \bra{e,m}\hat{\sigma}_+\ket{g,n} = \prescript{}{b}{\bra{m}} \hat{S}(-2r)\ket{n}_b \neq \delta_{m,n},
\end{equation}
where $\delta_{m,n}$ is the Kronecker delta.
This inner product is the same overlap in spatial wavefunctions between different mechanical states as discussed in Sec.\ref{sec:FrankCondon}, $\abs{\alpha_{mn}}^2 = \int_{-\infty}^{\infty} \psi^*(x,n,-1)\psi(x,m,1) dx$, and it only allows transitions that change the phonon occupation by an even number.

Going beyond the intuition, we calculate the transition rates that determine the qubit spectrum by writing down the time evolution of the system without decoherence under a coherent qubit drive (see Sec.\ref{sec:dynamics} for effects of decoherence),
\begin{equation}\label{eqn: drive Hamiltonian}
H_D(t) = \Omega_R \hat{\sigma}_x \cos{ \omega_d t},    
\end{equation}
where $\Omega_R$ is the Rabi rate due to the external drive, and $\omega_d$ is the drive frequency. Going into the interaction picture with, 
\begin{equation}\label{eqn: H0 interaction}
\begin{aligned}
    H_0 &= \hat{U}_\text{disp} \hat{S} \bigg(\mathcal{r}(\hat{\sigma}_z)\bigg) H_{\text{sq}} \hat{S}^\dagger \bigg(\mathcal{r}(\hat{\sigma}_z)\bigg) \hat{U}_\text{disp}^\dagger\\
    &= \sum_n \bigg( \omega_n^- \ket{g,n} \bra{g,n}
    + \omega_n^+ \ket{e,n} \bra{e,n} \bigg),
\end{aligned}
\end{equation}
the interaction Hamiltonian is given by,
\begin{equation}\label{eqn: interaction hamiltonian}
\begin{aligned}
    \mathcal{V}(t) &= e^{-iH_0t}H_D(t) e^{iH_0t} \\
    & = \frac{1}{2}\Omega_R \sum_{n,m} \bigg(
    \alpha_{mn}^* e^{-i\Delta_{mn}^dt}\ket{g,n}\bra{e,m} + \alpha_{mn} e^{i\Delta_{mn}^dt}\ket{e,m}\bra{g,n}\bigg),
\end{aligned}
\end{equation}
where $\alpha_{mn} = \prescript{}{b}{\bra{m}} \hat{S}(-2r)\ket{n}_b$ is the overlap between different motional states under a qubit flip (eqn.(\ref{eqn:mech state overlap})), and $\Delta_{mn}^d = \omega_m^+ - \omega_n^- - \omega_d$ is the detuning between the qubit drive and transition $\ket{g,n}\leftrightarrow \ket{e,m}$. 
Equation (\ref{eqn: interaction hamiltonian}) identifies the appearance of sideband transitions separated by $2\omega_m$.
Starting from an initial state $\ket{g,n}$, the qubit spectrum peaks at drive frequencies,
\begin{equation}
    \omega_d = \omega_q^n + 2l\omega_m  (k\in \mathbb{Z} ),
\end{equation}
with transition rates 
\begin{equation} \label{eqn:QuantumSBrate-degeneracy}
    \Omega_n^{2l} = \Omega_R\abs{\alpha_{n+2l,n}} = \Omega_R \abs{\prescript{}{b}{\bra{n+2l}}\hat{S}(-2r)\ket{n}_b}.
\end{equation}
In our measurement scheme, this transition rate is related to the spectroscopic qubit excitation probability $P_e \propto \Omega^2$\cite{Schuster2005AcField}.

\subsection{Effect of residual coupling}\label{sec:SigmaZ coupling}


Charge noise introduces a residual $\hat{\sigma}_z$ coupling between the qubit energy and mechanical oscillator position. Despite its small coupling rate $g_m^z \ll g_m$, the residual coupling is an important effect because it describes the coupling between qubit energy and mechanical position at first order; whereas the $\hat{\sigma}_x$ coupling describes the second order coupling of position to energy. 
In this section, we investigate its effect by directly diagonalizing the Hamiltonian with a residual $\hat{\sigma}_z$ coupling.

According to eqn.(\ref{eqn:coupling}), a charge offset in $n_g$ introduces a residual $\hat{\sigma}_z$ coupling, and the Hamiltonian becomes,

\begin{equation}\label{eqn: sigma_z hamiltonian}
    H_0^\text{rsd} = \omega_m \hat{a}^\dagger \hat{a} + \frac{1}{2}\omega_q \hat{\sigma}_z + g_m^x\hat{\sigma}_x(\hat{a}+\hat{a}^\dagger) +  g_m^z\hat{\sigma}_z(\hat{a}+\hat{a}^\dagger).
\end{equation}
For small residual coupling $g_m^z \ll g_m^x \approx g_m$, we apply consecutively the following unitary transformations to diagonalize this Hamiltonian,
\begin{equation}\label{H sigma_z diagonalize}
H_{\text{diag}} ^\text{rsd} = \hat{S}^\dagger \bigg(\mathcal{r}(\hat{\sigma}_z)\bigg)
\hat{\mathcal{D}}^\dagger \bigg( \alpha(\hat{\sigma}_z) \bigg)
\hat{R}^\dagger \bigg( \theta(\hat{X}) \bigg)
\hat{U}_\text{disp}^\dagger H_0 ^\text{rsd} \hat{U}_\text{disp}
\hat{\mathcal{D}} \bigg( \alpha(\hat{\sigma}_z) \bigg)
\hat{R} \bigg( \theta(\hat{X}) \bigg)
\hat{S} \bigg(\mathcal{r}(\hat{\sigma}_z)\bigg),  
\end{equation}
where
\begin{equation}\label{eqn:rotation}
    \hat{R} \bigg( \theta(\hat{X}) \bigg) = \text{exp}\bigg[ -i\frac{1}{2}\hat{\sigma}_y \arctan \bigg( \frac{\chi_m^z \hat{X}^2}{\omega_q/2 + g_m^z \hat{X} + \chi_m \hat{X}^2} \bigg) \bigg] \approx \hat{\mathbbm{1}}
\end{equation}
is the qubit rotation operator with $\hat{X} = \hat{a}+\hat{a}^\dagger$, $\chi_m^z = g_m^x g_m^z\big( \frac{1}{\Delta} + \frac{1}{\Sigma}\big) \ll \chi_m$, and
\begin{equation}\label{eqn:displace}
\begin{aligned}
    \hat{\mathcal{D}} \bigg( \alpha(\hat{\sigma}_z) \bigg) &= \text{exp}\bigg[\alpha(\hat{\sigma}_z)\hat{a}^\dagger - \alpha^* (\hat{\sigma}_z) \hat{a} \bigg],\\
    \alpha(\hat{\sigma}_z) &= -\frac{g_m^z\hat{\sigma}_z}{\omega_m + 4\chi_m\hat{\sigma}_z}\approx -\frac{g_m^z}{\omega_m}\hat{\sigma}_z \equiv \beta \hat{\sigma}_z
\end{aligned}
\end{equation}
is the qubit-state dependent displacement operator. Similar to eqn.(\ref{eqn:eigenstate main text}), the eigenstate of $H_0 ^\text{rsd}$ are given by,
\begin{equation}\label{eqn:eigenstate main text sigma_z}
\begin{aligned}
    \ket{g,n} &= \hat{S}(r\hat{\sigma}_z)\hat{\mathcal{D}}(\beta \hat{\sigma}_z) \ket{g}_b  \ket{n}_b = \ket{g}_b \hat{S}(-r)\hat{\mathcal{D}}(-\beta)\ket{n}_b,\\
    \ket{e,n} &= \hat{S}(r\hat{\sigma}_z)\hat{\mathcal{D}}(\beta \hat{\sigma}_z)\ket{e}_b  \ket{n}_b = \ket{e}_b \hat{S}(r)\hat{\mathcal{D}}(\beta)\ket{n}_b.
\end{aligned}    
\end{equation}
Consequently, the overlap between different mechanical states under a qubit excitation is given by,
\begin{equation}\label{eqn:mech state overlap sigma_z}
    \alpha_{mn} ^\text{rsd} = \bra{e,m}\hat{\sigma}_+\ket{g,n} = \prescript{}{b}{\bra{m}}\hat{\mathcal{D}}^\dagger(\beta) \hat{S}(-2r) \hat{\mathcal{D}}(-\beta)\ket{n}_b.
\end{equation}
Therefore, as the residual $\hat{\sigma}_z$ coupling breaks the symmetry in the system, a qubit excitation can connect all mechanical states. In terms of qubit spectroscopy, if we start from an initial state of $\ket{g,n}$, we expect qubit excitation peaks centered at drive frequencies 
\begin{equation}
    \omega_d = \omega_q^n + l\omega_m (l \in \mathbb{Z})
\end{equation}
separated by $\omega_m$, with transition rates 
\begin{equation}\label{eqn:QuantumSBrate}
    \Omega_n^l = \Omega_{R}\abs{\alpha_{n+l,n} ^\text{rsd}}.
\end{equation}

\section{Simulating Fock state response with a classical drive}\label{sec:qubit phonon responce}

To simulate the effect of motion, we apply a large ac gate-voltage at $\omega_m$ on the dc bias line, which is weakly coupled to the qubit. This drive behaves as a classical modulation of the qubit's gate-charge $n_g$. In Sec.\ref{sec:Principle classical model}, we justify this simulation, and describe the effects of the classical gate-charge modulation in terms of qubit spectroscopy. In Sec.\ref{sec:ClassicalVsQuantum} $\&$ \ref{sec:charge noise}, we show the qubit response to classical drive to be a good approximation of that to quantized motion in the mechanical oscillator. Finally in Sec.\ref{sec:map}, we show the procedure of converting the measured qubit response to a point-spread-function (PSF)\cite{Richardson1972LRreconstruction,PRLsupplementary} map used to reconstruct phonon distributions from qubit spectra. 

\subsection{Classical modulation on gate-charge}\label{sec:Principle classical model}

As discussed in Sec.\ref{sec:EquivalentCircuit}, the coupling originates from a oscillator-position-dependent gate-charge $n_g(x)$. If the motion is treated classically, it will sinusoidally modulate the gate-charge in time at the mechanical frequency. 
To experimentally simulate this effect, we turn off the qubit-mechanics coupling by setting $V_\text{dc} = 0$, and drive an ac voltage through the same dc bias line (Sec.\ref{sec:Full}) at $\omega_m$. This ac voltage is converted into a CPB gate-voltage through the asymmetry between $C_0$ and $C_c$, and results in a time dependent gate-charge 
\begin{equation}
    n_g(t) = \frac{1}{2} + \delta n_g + n_x \cos(\omega_m t),
\end{equation}
where $n_x$ is the modulation amplitude, and $\delta n_g$ is an offset from the charge degeneracy point.

From this modulation, we find the qubit spectrum to be altered in two distinct ways. First, the center qubit resonance is continuously shifted, $\delta \omega_q^x(\delta n_g, n_x)$, as a function of modulation amplitude (eqn.\ref{eqn:Stark shift charge noise}). Second, the modulation will cause the appearence of sideband peaks at large modulation amplitude. At the charge degeneracy point ($\delta n_g = 0$), those sidebands are separated by $2\omega_m$ with a transition rate given by Bessel functions of the first kind $\Omega_x^{2l} = \Omega_R J_l \big( \frac{\delta \omega_q^x(n_x)}{2\omega_m} \big)$ (eqn.\ref{eqn: classical interaction hamiltonian}). Alternatively, when an offset charge is present ($\delta n_g \neq 0$), those sidebands are separated by $\omega_m$ with a transition rate given by a product of Bessel functions (eqn.\ref{eqn: classical interaction Hamiltonian charge noise}). 

We start at the charge degeneracy point. Expanding around small modulation amplitude, the qubit frequency (eqn.(\ref{eqn:qubit frequency})) is
\begin{equation}\label{eqn:qubit frequency classical modulation}
    \begin{aligned}
        \omega_q(t) &= \sqrt{E_J^2+(4E_c)^2(1-2n_g(t))^2}\\
        & = E_J + 2\frac{(4E_c)^2}{E_J} n_x^2 \cos^2(\omega_m t) + \mathcal{O}(n_x^4)\\
        & \approx E_J + \frac{(4E_c)^2}{E_J}n_x^2 + \frac{(4E_c)^2}{E_J}n_x^2 \cos(2\omega_m t).
    \end{aligned}
\end{equation}
Here, the first term is the bare qubit frequency at the charge degeneracy point, $\omega_q^b = E_J$. The second term corresponds to a static frequency shift of the qubit that is proportional to the energy in the classical drive, analogous to the ac Stark shift. 
The last term in eqn.(\ref{eqn:qubit frequency classical modulation}) is a frequency modulation of the qubit frequency at $2\omega_m$. To understand its effect, we write down the time evolution of the system without decoherence under a coherent qubit drive given by eqn.(\ref{eqn: drive Hamiltonian}). Here, the bare Hamiltonian is $H_0^\text{c} = \frac{1}{2}\omega_q(t)\hat{\sigma}_z$, and its time evolution operator is,
\begin{equation}\label{eqn: classical H0 time evolution}
\begin{aligned}
    \hat{U}_0^c(t) &= \hat{\mathcal{T}}\bigg( \text{exp} \left[-i\int_0^t H_0^\text{c} (\tau) d\tau \right] \bigg)
    = \text{exp} \left[ -\frac{i}{2} \phi(t)  \hat{\sigma}_z \right],    
\end{aligned}
\end{equation}
where $\hat{\mathcal{T}}$ is the time ordering operator, and
\begin{equation}\label{eqn:time dependent phase}
    \phi(t) = \int_0^t \omega_q(\tau) d\tau = \omega_q^x(n_x) t + \frac{ \delta \omega_q^x}{2 \omega_m} \sin(2\omega_m t),
\end{equation}
where 
\begin{equation}\label{eqn:stark shift classical}
   \omega_q^x(n_x) = \omega_q^b + \frac{(4E_c)^2}{E_J} n_x^2 
\end{equation}
is the drive power dependent qubit mean frequency, and $\delta \omega_q^x (n_x) = \omega_q^x(n_x) -\omega_q^b$ is the qubit-frequency shift because of the classical modulation. Therefore, invoking the Jacobi-Anger expansion, the time evolution for the coherent drive in the interaction picture is,
\begin{equation}\label{eqn: classical interaction hamiltonian}
\begin{aligned}
    \mathcal{V}^c(t) &= \hat{U}_0^c {}^\dagger(t) H_R(t) \hat{U}_0^c(t) = \frac{1}{2}\Omega_R \bigg( \hat{\sigma}_+ e^{i \left[ \phi(t)-\omega_d t \right]} + \hat{\sigma}_- e^{-i\left[\phi(t)-\omega_d t\right]} \bigg)\\
    & = \sum_{l = -\infty}^\infty \frac{1}{2} \Omega_R J_l \bigg( \frac{\delta \omega_q^x(n_x)}{2\omega_m} \bigg) \bigg[ \hat{\sigma}_+ e^{i\Delta_{l,x}^d(n_x) t} + \hat{\sigma}_- e^{-i\Delta_{l,x}^d(n_x) t} \bigg]
\end{aligned}
\end{equation}
where $J_l(z)$ is the $l$-th order Bessel function of the first kind, and
\begin{equation}
    \Delta_{l,x}^d(n_x) = \omega_q^x(n_x) + 2l\omega_m -\omega_d.
\end{equation}
Similar to eqn.(\ref{eqn: interaction hamiltonian}), eqn.(\ref{eqn: classical interaction hamiltonian}) identifies the appearance of sideband transitions at frequencies $\omega_q^x(n_x)$, separated by $2\omega_m$. Coherently driving at one of these resonant frequencies, the transition rate is given by 
\begin{equation}\label{eqn:ClassicalSBRateDegeneracy}
    \Omega_x^{2l} = \Omega_R J_l \bigg( \frac{\delta \omega_q^x(n_x)}{2\omega_m} \bigg).
\end{equation}

A similar calculation with a static offset $\delta n_g$ from the charge degeneracy point recovers the sideband transitions separated by odd integer multiples of $\omega_m$, and captures the classical effect due to residual $\hat{\sigma}_z$ coupling (Sec.\ref{sec:SigmaZ coupling}). Writing the time dependent gate-charge as $n_g(t) = \frac{1}{2} + \delta n_g + n_x \cos{(\omega_m t)}$, eqn.(\ref{eqn:qubit frequency classical modulation}) becomes
\begin{equation}\label{eqn: qubit frequency classical modulation offset}
    \omega_q(t) \approx \omega_q^b(\delta n_g) + \delta \omega_q^x(\delta n_g, n_x) + \delta \omega_q^x(\delta n_g, n_x) \cos(2\omega_m t) + \frac{(8E_c)^2}{\omega_q(\delta n_g)}n_x \delta n_g  \cos{(\omega_m t)},
\end{equation}
where 
\begin{equation}\label{eqn:bare qubit frequency charge noise}
    \omega_q^b(\delta n_g) = \sqrt{E_J^2 + (8E_c \delta n_g)^2},
\end{equation}
is the bare qubit frequency at charge offset $\delta n_g$, and
\begin{equation}\label{eqn:Stark shift charge noise}
    \delta \omega_q^x(\delta n_g, n_x) = \frac{(4E_c E_J)^2}{\left( \omega_q^b(\delta n_g) \right)^3}  n_x^2
\end{equation}
is the drive-induced qubit shift. Note that at charge degeneracy, $\delta n_g = 0$, $\omega_q^b(0)=E_J$, and we recover eqn.(\ref{eqn:stark shift classical}). We can also find the time evolution for the coherent drive in the interaction picture to be
\begin{equation}\label{eqn: classical interaction Hamiltonian charge noise}
    \mathcal{V}^c(t) = \sum_{a,b = -\infty}^\infty  \frac{1}{2} \Omega_R J_a \bigg( \frac{\delta \omega_q^x (\delta n_g, n_x)}{2\omega_m} \bigg)J_b \bigg( \frac{(8E_c)^2}{\omega_m \omega_q^b (\delta n_g)} n_x \delta n_g \bigg) \bigg[ \hat{\sigma}_+ e^{i\Delta_{a,b,x}^d(\delta n_g,n_x) t} + \hat{\sigma}_- e^{-i\Delta_{a,b,x}^d(\delta n_g,n_x) t} \bigg],
\end{equation}
where 
\begin{equation}
    \Delta_{a,b,x}^d(\delta n_g,n_x) = \omega_q^x(\delta n_g,n_x) + (2a+b)\omega_m -\omega_d.
\end{equation}
This identifies the appearence of sideband transitions separated by $\omega_m$. Coherently driving at one of these frequencies, the transition rate is given by 
\begin{equation}\label{eqn:ClassicalSBRate}
    \Omega_x^l = \Omega_R \displaystyle \sum_{2a+b=l} J_a \big( \frac{\delta \omega_q^x (\delta n_g, n_x)}{2\omega_m} \big)J_b \big( \frac{(8E_c)^2}{\omega_m \omega_q^b (\delta n_g)} n_x \delta n_g \big).
\end{equation}

\subsection{Classical model vs. quantum theory}\label{sec:ClassicalVsQuantum}

In general, the qubit response to quantized motion is different from the qubit response to classical modulation. Despite the common qubit-frequency shift proportional to modulation power and the same resonant conditions under a coherent drive, the difference between the two lies in their different transition rates. For simplicity, here we only discuss the case where the static gate-charge is at the charge degeneracy point ($\delta n = 0$).
For a classical modulation around this point, the transition rates are symmetric around the center qubit peak, $\Omega_x^{2l} = \Omega_x^{-2l}$ (eqn.(\ref{eqn:ClassicalSBRateDegeneracy})), and the qubit spectroscopy is therefore also symmetric. With quantized motion, however, the qubit spectroscopy is asymmetric around the phonon-conserving peak. 
A simple example is the case of an initial state $\ket{g,0}$. While phonons can be added into the mechanical oscillator using blue sideband transitions, they cannot be extracted from the mechanical ground state using red sideband transitions.

In this experiment, however, the qubit response to quantized motion is well approximated by the qubit response to classical modulation. When the sideband transitions become prominent features in qubit spectroscopy at $n \gtrsim 100$ (because $\chi_m/\omega_m \approx 0.01$), the asymmetry between the blue and red sideband transitions is small enough to be neglected. Conversely, when the asymmetry is strong at small phonon numbers, the qubit spectroscopy is dominated by the phonon conserving transition and the sidebands can be all together ignored.
To provide an intuition for this, we calculate the asymmetry by expanding the sideband rate $\Omega_n^{\pm2}$ connecting states $\ket{g,n}$ and $\ket{g,n\pm2}$ to first order in $\chi_m/\omega_m$ (eqn.(\ref{eqn:QuantumSBrate-degeneracy})),
\begin{equation}
    \frac{\Omega_n^{-2}}{\Omega_n^{2}} \approx \sqrt{\frac{n(n-1)}{(n+1)(n+2)}} \approx 1- \frac{2}{n} + \mathcal{O}(\frac{1}{n^2}).
\end{equation}
At $n=100$ phonons, this corresponds to a $\sim 2\%$ asymmetry between sideband rates and a $\sim 1\%$ deviation from the classical case. Indeed a negligible amount compared to the measurement noise in the qubit spectrum. 

\begin{figure} [h!]
    \centering
    \includegraphics{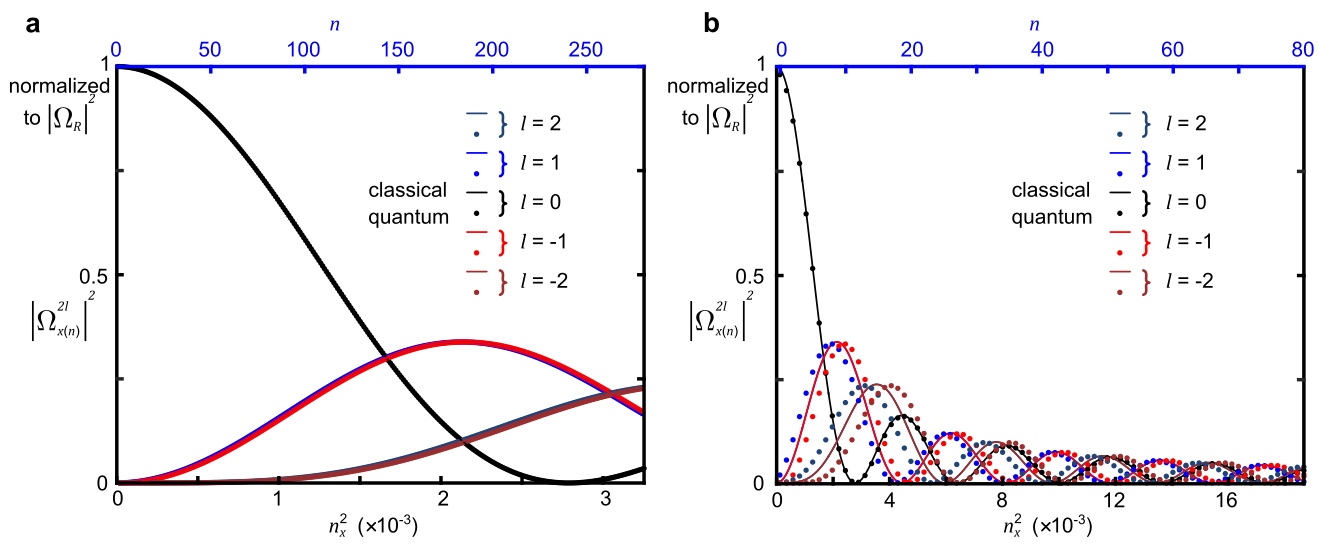}
    \caption{\textbf{Transition rates under classical modulation vs. quantized motion}\\ 
    The solid lines correspond to transition rates squared $\abs{\Omega_x^{2l}}^2$ as functions of the $n_g$ modulation amplitude squared $n_x^2$ (bottom-axis) according to eqn.(\ref{eqn: classical interaction hamiltonian}). The dots correspond to transition rates squared $\abs{\Omega_n^{2l}}^2$ for a given initial phonon number $n$ (top-axis) according to eqn.(\ref{eqn: interaction hamiltonian}). For a given squeezing amplitude $\big($\textbf{a,} $2r \approx \chi_m / \omega_m = 0.01$ and \textbf{b,} $2r = 0.2 \big)$, $n_x^2$ is related to $n$ by enforcing the same qubit-frequency shift according to eqn.(\ref{eqn:ng to nph}). The quality of approximation using classical gate-charge modulation degrades for larger squeezing amplitude.}
    \label{fig:classical vs. quantum}
\end{figure}

A more rigorous numerical comparison can be made between the transition rates under classical and quantum modulation using eqn.(\ref{eqn:ClassicalSBRateDegeneracy}) and eqn.(\ref{eqn:QuantumSBrate-degeneracy}). 
To perform this comparison, we first relate the motional quanta to the equivalent charge modulation amplitude through the qubit-frequency shift. At charge degeneracy, using eqn.(\ref{eqn:stark shift quantum}) and eqn.(\ref{eqn:stark shift classical}) to shift the qubit frequency by the same amount, we find 
\begin{equation}\label{eqn:ng to nph}
    n + \frac{1}{2} = \frac{(4E_c)^2}{2\chi_m E_J}n_x^2,
\end{equation}
where $n$ is the motional quanta, and $n_x$ is the modulation amplitude of gate-charge.
\textbf{Fig.}\ref{fig:classical vs. quantum} shows numerical comparisons of the transition rate under classical and quantum theory. Although different in general (\textbf{Fig.}\ref{fig:classical vs. quantum}(b)), for us at $2r \approx \chi_m / \omega_m \approx 0.01$, \textbf{Fig.}\ref{fig:classical vs. quantum}(a) shows the simulation with classical modulation to be a good approximation for qubit response to quantized motion in the mechanical oscillator. 

In the above discussion, we only considered the case at the degeneracy point, $\delta n_g = 0$. Nevertheless, we can follow the same calculation for Fig.~\ref{fig:classical vs. quantum map} and compare the transition rates between a classical modulation (eqn.(\ref{eqn:ClassicalSBRate})) and quantized motion (eqn.(\ref{eqn:QuantumSBrate})) with values of $\delta n_g \neq 0$. 
For all reasonable values of $\delta n$ (within $\pm$3 standard deviations for the gate-charge offset $\abs{\delta n} \leq 3\sigma_c$, see next section), we find the simulation with classical modulation to remain a good approximation for qubit response to quantized motion.

\subsection{Charge noise}\label{sec:charge noise}
 
In this section, we explore the effects of a $1/f$ gate-charge noise on the measured qubit spectrum. While charge noise at $\omega_m$ introduces an uncertainty in $n_x^2$ and thus in extracted phonon number, charge noise with frequency component much less than $\omega_m$ leads to an asymmetric qubit lineshape. We use this asymmetry to extract a charge noise intensity consistent with typical reported values. 
Based on this extracted noise intensity, we calculate the qubit spectrum as shown in \textbf{Fig.}\ref{fig:classical vs. quantum map}(c). This calculation predicts the odd-order sideband peaks and agrees well with the measurement shown in \textbf{Fig.}\ref{fig:classical vs. quantum map}(b). 

To understand the effects of charge noise, we model it as a sum of sinusoidal signals with frequencies $\omega_i$ and random phases $\phi_i$ uniformly distributed within range $[0,2\pi]$,
\begin{equation}
    \sum_i n_i \cos{(\omega_i t + \phi_i)},
\end{equation}
where $n_i$ is the noise amplitude. The overall time dependence of the gate-charge centered around the degeneracy point is therefor $n_g(t) = \frac{1}{2}+ n_x\cos{(\omega_m t)} + \sum_i n_i \cos{(\omega_i t + \phi_i)}$. Following eqn.(\ref{eqn:qubit frequency classical modulation}), and treating both $n_x$ and $n_i$ as small parameters, the time dependent qubit frequency is
\begin{equation}\label{eqn:classical modulation charge noise}
\begin{aligned}
    \omega_q(t) =  \omega_q^b & + \frac{(4Ec)^2}{E_J}n_x^2 + \frac{(4Ec)^2}{E_J}n_x^2\cos{(2\omega_m t)}\\
    & + \frac{(8E_c)^2}{E_J} \sum_{i,j}n_i n_j \cos{(\omega_i t+\phi_i)}\cos{(\omega_j t+\phi_j)} \\
    & + \frac{(8E_c)^2}{E_J} \sum_i n_x n_i \cos{(\omega_m t)}\cos{(\omega_i t+\phi_i)} \\
    & + \mathcal{O}(n_x^4) + \mathcal{O}(n_i^4) + \mathcal{O}(n_i^2 n_x^2).
\end{aligned}
\end{equation}
Here, the first line is identical to eqn.(\ref{eqn:qubit frequency classical modulation}) and describes the effect of the explicit classical drive on the gate.

\begin{figure} [h!]
    \centering
    \includegraphics{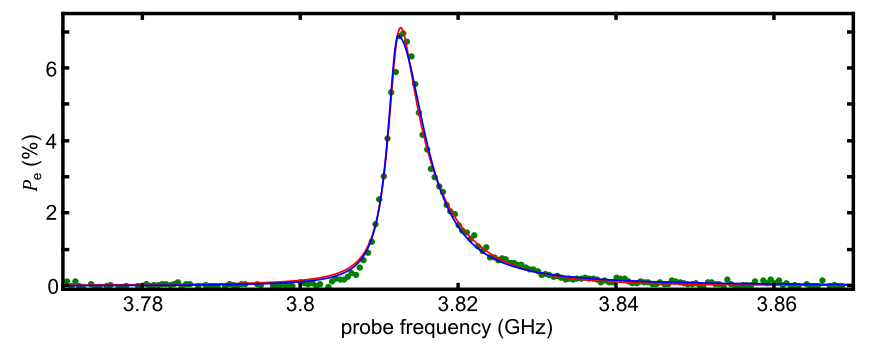}
    \caption{\textbf{Asymmetric qubit lineshape due to charge noise}\\
    At $V_\text{dc} = 0$V, and $n_x = 0$, we measure the qubit spectrum (green dots).
    Because of the highly averaged measurement protocol, the $1/f$ charge noise manifests as an asymmetric qubit lineshape. Modeling the charge noise as a Gaussian random variable, we can fit (red) the qubit lineshape with eqn.(\ref{eqn:charge noise qubit spectrum}) to find the standard deviation in gate-charge, $\sigma_c = 0.0071$ (2$e$). Alternatively, we can fit the qubit lineshape with skewed Lorentzians (blue) without assumptions of charge noise distribution according to eqn.(\ref{eqn:Lorentzian fit}). We use this skewed Lorentzian fit to process the measured qubit spectra under classical charge modulation in Sec.\ref{sec:map}.
    }
    \label{fig:charge noise}
\end{figure}

An asymmetric qubit lineshape arises from the dispersion relation of the qubit frequency around the charge degeneracy point (eqn.(\ref{eqn:qubit frequency})). This corresponds to the second line of eqn.(\ref{eqn:classical modulation charge noise}), which describes the qubit response without the classical ac drive. Because measurements are averaged over many realizations, this effect is best understood by treating the incoherent charge noise as a set of stationary charge offsets $\{ \delta n_g \}$ away from $n_g = 1/2$. Assuming a Gaussian process for this offset with mean $\langle \delta n_g \rangle = 0$ and standard deviation $\sigma_c$, the averaged qubit spectrum with $n_x = 0$ is a sum of qubit spectra weighted by the corresponding offset probability,
\begin{equation}\label{eqn:charge noise qubit spectrum}
    P_e(\omega) = \int_{d \delta n_g} \frac{1}{\sqrt{2\pi \sigma_c^2}} \text{exp}\big( -\frac{\delta n_g^2}{2\sigma_c^2} \big) \left[ \frac{(A \Gamma_\text{intrinsic}/2)^2}{\left( \omega - \omega_q^b(\delta n_g) \right)^2 + (\Gamma_\text{intrinsic}/2)^2 (1+A^2)} \right] d \delta n_g,
\end{equation}
where $\Gamma_\text{intrinsic}$ is the qubit linewidth without charge noise, $A = \Omega_R\sqrt{\frac{2}{\Gamma_1 \Gamma_\text{intrinsic}}}$ is the reduced Rabi rate, and $\omega_q^b(\delta n_g) = \sqrt{E_J^2 + (8E_c\delta n_g)^2}$ is the bare qubit resonance given $\delta n_g$. As shown in \textbf{Fig.}\ref{fig:charge noise}, with $\Gamma_\text{intrinsic}$, $A$, $E_J$ and $\sigma_c$ being free parameters, we fit the bare qubit lineshape ($V_\text{dc} = 0$, and $n_x = 0$) to find $\sigma_c = 0.0071$ (2$e$). Given the measurement protocol, this value corresponds to a $1/f$ charge noise intensity of $1.03\times 10^{-3} e/\sqrt{\text{Hz}}$ at 10~Hz, consistent with the typical charge noise intensity of $10^{-3} - 10^{-4}$ $e/\sqrt{\text{Hz}}$ at 10~Hz\cite{Zorin1996BackgroundDevices}.

Additional to the asymmetric lineshape, charge noise with frequency components close to $\omega_m$ also interacts with the classical modulation to introduce an uncertainty in the extracted phonon number, corresponding to the third line of eqn.(\ref{eqn:classical modulation charge noise}). Using the charge noise intensity found above, we find the uncertainty in phonon number to be negligible even at $n \sim 300$.

Thus, we find the simulation with classical modulation to remain faithful in the presence of charge because the noise-induced phonon number uncertainty is negligible and the noise-induced asymmetry is fully captured by the highly averaged qubit measurement.

Finally, we provide a model that predicts the qubit spectrum under an explicit classical charge modulation of amplitude $n_x$ in the presence of charge noise.
We assume a Gaussian charge noise process and a field-strength of $\epsilon$ for the spectroscopic drive.
For a given gate-charge offset $\delta n_g$, the expected qubit spectrum is given by a sum of Lorentzians,
\begin{equation}\label{eqn:spectrum with charge offset and modulation}
    P_e^{\delta n_g}(n_x,\omega) = \sum_l\frac{1}{2}\frac{(A_l(\omega) \Gamma_\text{intrinsic} / 2)^2}{(\omega - \omega_l)^2 + (\Gamma_\text{intrinsic})^2 (1+A_l^2(\omega))},
\end{equation}
where $\omega_l= \omega_q^b(\delta n_g) + \delta \omega_q^x (\delta n_g,n_x) + l\omega_m$ is the resonance frequency of the $l$-th order sideband transition, and
\begin{equation}\label{eqn:reduced rabi rate sideband}
\begin{aligned}
    A_l(\omega) &= \Omega_x^l(\omega)\sqrt{\frac{2}{\Gamma_1 \Gamma_\text{intrinsic}}} \\
    &= \Omega_R(\omega) \sqrt{\frac{2}{\Gamma_1 \Gamma_\text{intrinsic}}} \sum_{2a+b=l} J_a \big( \frac{\delta \omega_q^x (\delta n_g, n_x)}{2\omega_m} \big)J_b \big( \frac{(8E_c)^2}{\omega_m \omega_q^b (\delta n_g)} n_x \delta n_g \big)
\end{aligned}
\end{equation}
is its reduced transition rate with $\Omega_R(\omega) = 2 g_c \epsilon/(\omega - \omega_c)$ being the drive Rabi rate. Similar to eqn.(\ref{eqn:charge noise qubit spectrum}), we model the charge noise as a Gaussian process with $\langle \delta n_g \rangle= 0$ and standard deviation $\sigma_c$ to find the overall qubit spectrum
\begin{equation}\label{eqn:charge noise qubit spectrum with modulation}
    P_e(n_x,\omega) = \int_{d\delta n_g} \frac{1}{\sqrt{2 \pi \sigma_c^2}} \text{exp}\left(-\frac{{\delta n_g}^2}{2\sigma_c^2} \right) P_e^{\delta n_g}(n_x,\omega) d \delta n_g. 
\end{equation}
\textbf{Fig.}\ref{fig:classical vs. quantum map}(c) shows the numerical result of this model where $\Gamma_\text{intrinsic}$, and $\sigma_c = 0.0071 (2e)$ are both extracted from \textbf{Fig.}\ref{fig:charge noise}. Only the drive strength $\epsilon$ is left free to match the maximum peak-height in the map of \textbf{Fig.}\ref{fig:classical vs. quantum map}(b). In this figure, we observe the odd order sideband transitions discussed in Sec.\ref{sec:SigmaZ coupling} \& \ref{sec:Principle classical model}.
The good agreement between the experiment shown in \textbf{Fig.}\ref{fig:classical vs. quantum map}(b) and the theory shown in \textbf{Fig.}\ref{fig:classical vs. quantum map}(c) allows us to attribute charge noise as the main source of residual $\hat{\sigma}_z$ coupling.

\subsection{Using the simulation}\label{sec:map}

\begin{figure} [h!]
    \centering
    \includegraphics{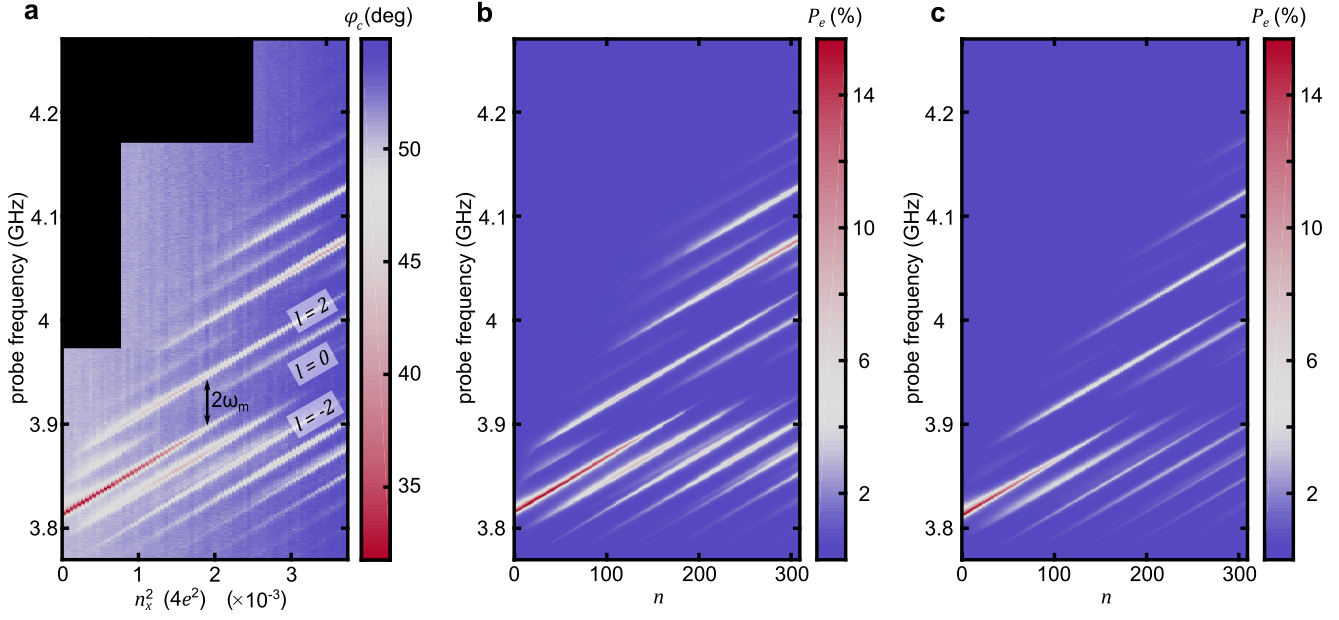}
    \caption{\textbf{Qubit response to classical modulation vs. quantized motion}\\
    \textbf{a,} We sweep the frequency of a weak qubit-excitation probe signal (y-axis) and measure the cavity transmission phase $\varphi_c$ (color-scale) as a function of the classical modulation amplitude squared $n_x^2$ (x-axis). 
    The qubit excitations appear as dips in $\varphi_c$.
    The background phase increases at larger $n_x$ because the gate-charge modulation around $n_g = 1/2$ rectifies the cavity phase response of \textbf{Fig.}\ref{fig:CPB}(b). Data is not collected in the black region because no feature is expected.
    \textbf{b,} We convert the measured phase response (\textbf{a}) to the qubit spectra (color-scale vs. y-axis) at particular mechanical Fock states (x-axis). The cavity transmission phase is fitted (eqn.\ref{eqn:Lorentzian fit}), converted to the probability of exciting the qubit, and interpolated at appropriate qubit Stark shifts (eqn.\ref{eqn:stark shift quantum}). We use this map to extract phonon distributions.
    \textbf{c,} We calculate the expected qubit spectra using eqn.(\ref{eqn:charge noise qubit spectrum with modulation}) and assuming a Gaussian gate-charge distribution of mean $\langle n_g \rangle = 0.5$ and standard deviation $\sigma_c = 0.0071$.}
    \label{fig:classical vs. quantum map}
\end{figure}

We convert the experimentally measured qubit response to classical modulation (\textbf{Fig.}\ref{fig:classical vs. quantum map}(a)) to a 
map of qubit response to quantized motion of quantized motion in two steps: 

(1) We connect the measured cavity transmission phase to qubit excitation probability $P_e(\omega)$ using eqn.(\ref{eqn:Pe}). Instead of directly using the measured phase which contains statistical noise, we instead fit the measured spectrum using a sum of skewed Lorentzians where each Lorentzian describes one sideband peak or the center qubit peak:
\begin{equation} \label{eqn:Lorentzian fit}
     \varphi_c(\omega) = \sum_l\frac{1}{2} \frac{\left( A_l\Gamma_l/2 \right)^2} {\left( \frac{\omega-\omega_l}{ 1+\text{L Sgn}(\omega - \omega_l)} \right)^2 + \left( \Gamma_l/2 \right)^2 \left(1+A_l^2 \right)},
\end{equation}
where $\omega_l$ is the resonances of the system under the spectroscopic drive, $A_l$ describes the corresponding transition rate $\abs{\Omega_x^l}^2$, and $L$ captures the asymmetry due to charge noise\cite{PRLsupplementary}. In this fit, we also remove the background phase present in \textbf{Fig.}\ref{fig:classical vs. quantum map}(a), which increases at larger modulation amplitude. This background phase arises from time-averaging the cavity phase response of \textbf{Fig.}\ref{fig:CPB}(b) to a classical gate-charge modulation around $n_g = 1/2$.

(2) We find the qubit response at integer phonon numbers by interpolating the fit parameters extracted from step (1). 

We thus create a new map, as shown in \textbf{Fig.}\ref{fig:classical vs. quantum map}(b), which we take as the qubit response to mechanical Fock states. As discussed in the main text, we perform deconvolution procedures and least-squared fits to understand phonon distribution. Details of those techniques can be found in reference \cite{PRLsupplementary}, replacing the PSF ($\Pi_{ni}$) with the map in \textbf{Fig.}\ref{fig:classical vs. quantum map}(b).

Finally, two technical details need to be considered to properly reconstruct the mechanical phonon distribution using \textbf{Fig.}\ref{fig:classical vs. quantum map}(b). First, because qubit lineshape is strongly dependent on the spectroscopic drive strength\cite{Schuster2005AcField}, we ensure the drive power is the same for all measurements. 
Second, the bare qubit frequency is different between experiments performed at $V_\text{dc} = 0$~V and $V_\text{dc} = 6$~V, with a difference of $3-3.5$~MHz. 
We attribute this to both a change in qubit charging energy due to the dc voltage\cite{PRLsupplementary} and a change in Josephson energy due to variations in the local flux.
To correct for this error, we measure the qubit spectrum with the mechanical oscillator in a thermal state before each experiment of interest, and perform a least-squared fit to extract the bare qubit frequency (see Sec.\ref{sec:coherent drive}).

\section{Measurement backaction}\label{sec:backaction}

We infer the phonon distribution from the qubit spectrum, but this inferences requires care because the qubit state is not a quantum-nondemolition measurement of the phonon number. On the contrary, the sidebands seen in the qubit spectrum correspond to the creation or annihilation of phonons. Thus, to ensure an accurate inference of the phonon distribution, we need to limit the measurement-induced change in phonon distribution to a negligible amount. To this end, we employ a short spectroscopic drive duration, and demonstrate its perturbation on phonon distribution is small beyond detection.

\begin{figure} [h!]
    \centering
    \includegraphics{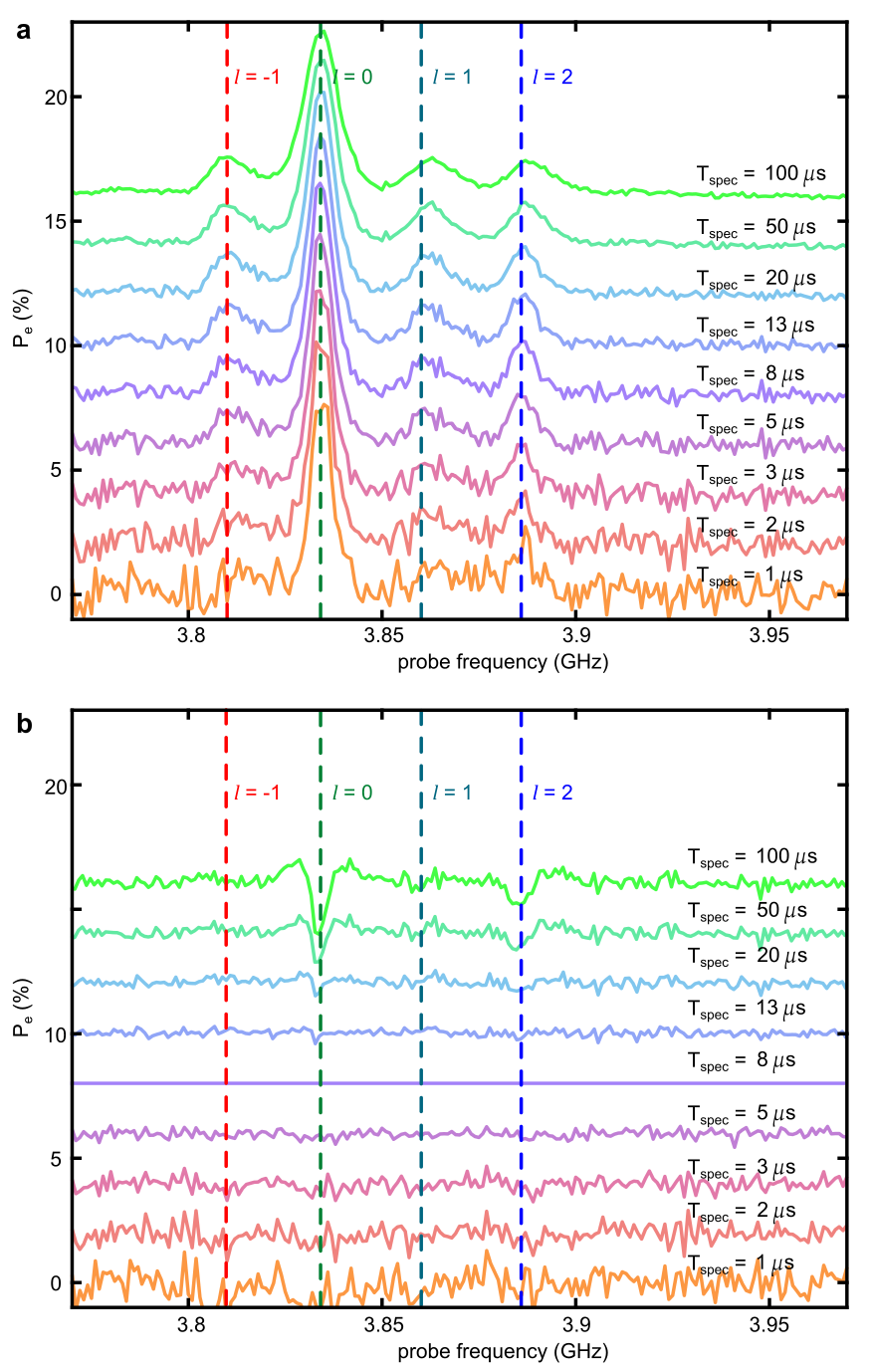}
    \caption{\textbf{Measurement backaction}\\
    \textbf{a,} We measure the qubit spectrum with different durations $T_{\text{spec}}$ of the spectroscopy tone. The oscillator has been prepared into a sub-Poissonian state with $\langle n \rangle = 43$. The trace with $T_{\text{spec}}=8~\mu$s is identical to Fig.4(a) of the main text. An incremental offset of $2\%$ is added for each trace. The vertical dashed lines correspond to the positions of the qubit excitation peaks at $T_{\text{spec}}=8~\mu$s. With increasing $T_{\text{spec}}$, the $l=-1$ peak shifts toward a lower frequency because of sideband cooling, and the $l=\{1,2\}$ peaks shift toward higher frequencies because of sideband heating, demonstrating measurement backaction. With increasing $T_{\text{spec}}$, we also observe a broadening of the center qubit peak caused by the damping of the mechanical oscillator during measurement.    
    \textbf{b,} We subtract the $T_{\text{spec}} = 8~\mu$s qubit spectrum from each qubit spectrum in \textbf{a}. Structure can be observed in the traces with $T_{\text{spec}} \geqslant 13~\mu$s at the positions of the dashed lines.}
    \label{fig:backaction}
\end{figure}

For the core result of this paper (Fig.4 of the main text), we find the appropriate duration of $T_{\text{spec}} = 8~\mu$s experimentally as shown in \textbf{Fig.}\ref{fig:backaction}. After preparing the sub-Poissonian state as described in the main text, we drive the qubit for different $T_{\text{spec}}$ and measure its spectrum. When we increase $T_{\text{spec}}$ past $13~\mu$s, we start to observe changes in the qubit spectrum associated with changing phonon distribution. Therefore, we use $T_{\text{spec}} = 8~\mu$s for Fig.4(a) of the main text. However, we note that $T_{\text{spec}} = 8~\mu$s is not a universal condition for avoiding measurement backaction. Instead, it is influenced by both measurement noise and the phonon distribution being measured.

In experiments where the phonon distribution is already broad, and where we are interested in detecting how the distribution changes as function of other parameters (Fig. 2 and 3 in the main text),  we use longer spectroscopy pulses of $100~\mu$s. This improves the duty cycle of measurements, which use 8~ms repetition rates to allow for the mechanical oscillator to re-equilibrate, at the cost of a small reduction in the fidelity of the extracted phonon distributions. 

\section{Coherent mechanical displacement}\label{sec:coherent drive}

Coherent mechanical motion is driven resonantly with the product of the dc voltage $V_\text{dc}$ and an ac voltage applied on the same electrode. Because the mechanical oscillator is originally in thermal equilibrium with the dilution fridge, the mechanical states we prepare this way are coherently displaced thermal states, with phonon distribution given by\cite{deOliveira1990PropertiesStates},
\begin{equation}\label{eqn:displaced thermal distribution}
    P(n,n_\text{th},n_\text{disp})= \sum_{m=0}^\infty \frac{n_{\text{th}}^m}{\left(1+n_{\text{th}} \right)^{m+1}} e^{-n_{\text{disp}}} \frac{\text{min}\{ m,n \}!}{\text{max}\{ m,n \}!} n_{\text{disp}}^{\abs{n-m}} \left[L_{\text{min}\{ m,n \}}^{\abs{n-m}}(n_{\text{disp}}) \right]^2,   
\end{equation}
where $L_n^l(x)$ is the associated Laguerre polynomial, $n_{\text{th}}$ is the thermal occupation initially in the mechanical oscillator, and $n_{\text{disp}} = \abs{\alpha}^2$ is the mean phonon displacement due to the coherent drive. This equation allows us to perform least-squared fit on the measured displaced thermal states, as shown in Fig.2(d) of the main text.

\begin{figure} [h!]
    \centering
    \includegraphics{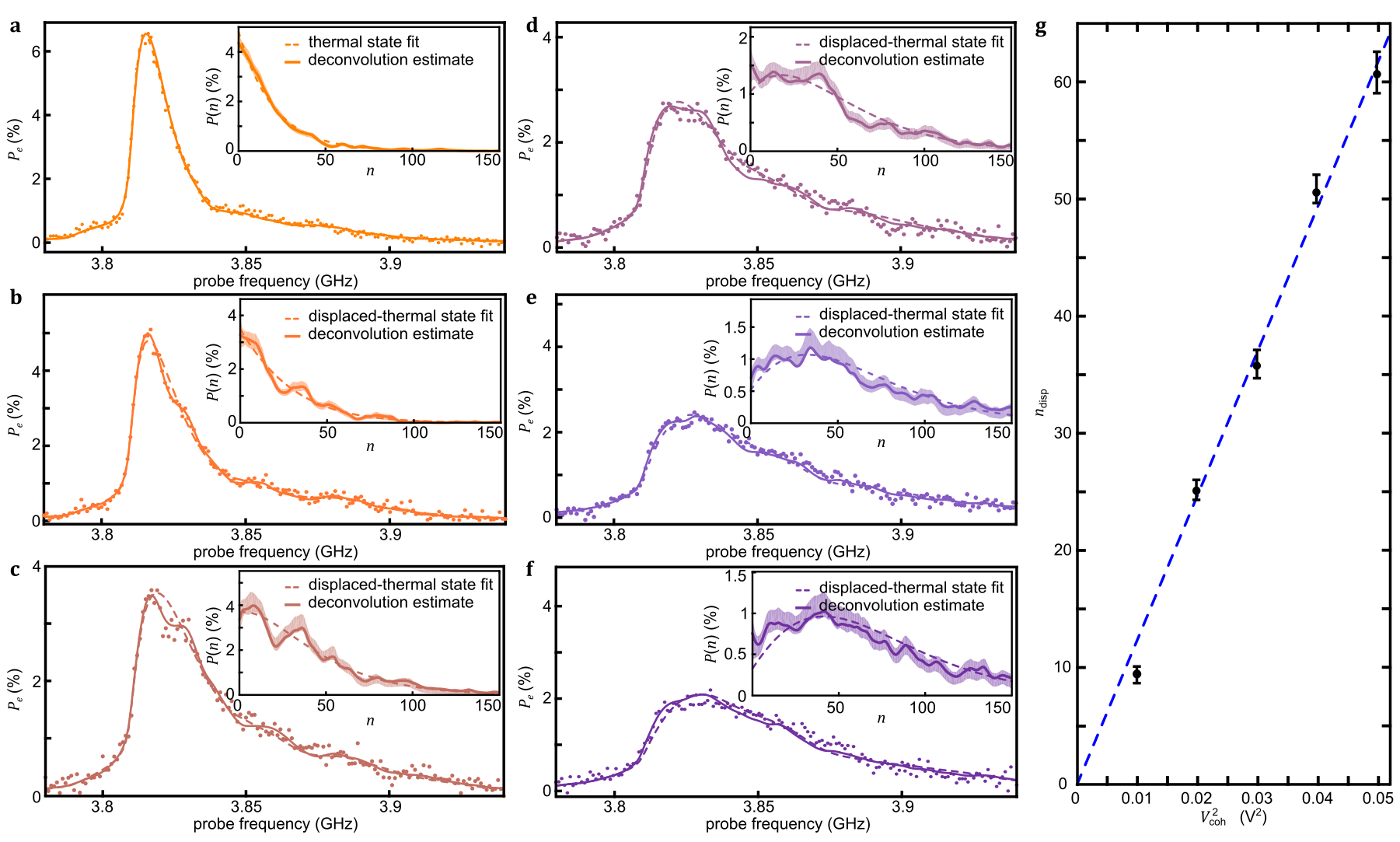}
    \caption{\textbf{Thermal and displaced thermal states}\\
    For different $V_{\text{coh}}$, we measure the qubit spectrum and extract the mechanical phonon distribution (inset). Phonon distributions extracted from deconvolution (solid) and their $90\%$ confidence interval (shaded) are plotted alongside the result of least-squared fit (dashed).
    \textbf{a,} With $V_{\text{coh}} = 0$~V, the mechanical oscillator is in a thermal state. A least-squared fit to the thermal distribution finds $n_{\text{th}} = 17.7$. This is identical to Fig.2(c) in the main text. 
    \textbf{b-f,} By applying an ac-drive with $25$~MHz frequency and amplitude $V_{\text{coh}}$ (specified at the generator output), we coherently displace the thermal mechanical state by a mean phonon displacement of $n_{\text{disp}}$. Keeping the extracted $n_{\text{th}}$ from \textbf{a} constant, we perform least-squared fits assuming displaced thermal distributions to extract $n_{\text{disp}}$. The extracted parameters are: \textbf{b,} $V_{\text{coh}} = 100.5$~mV and $n_{\text{dsip}} = 8.0$; \textbf{c,} $V_{\text{coh}} = 142$~mV and $n_{\text{dsip}} = 20.0$; \textbf{d,} $V_{\text{coh}} = 174$~mV and $n_{\text{dsip}} = 30.6$; \textbf{e,} $V_{\text{coh}} = 201$~mV and $n_{\text{dsip}} = 43.5$ (identical to Fig.2(d) of the main text); and \textbf{f,} $V_{\text{coh}} = 225$~mV and $n_{\text{dsip}} = 51.9$.
    \textbf{g,} We plot the extracted $n_{\text{dsip}}$ as a function of $V_{\text{coh}}^2$. The error bars correspond to $90\%$ confidence intervals of $n_{\text{dsip}}$ found through non-parametric bootstrapping. Dashed blue line is a linear fit that goes through the origin.}
    \label{fig:coherent displacement}
\end{figure}

Figure \ref{fig:coherent displacement}, shows more measured displaced thermal states. Similar to Fig.2 in the main text, we plot the measured qubit spectrum along with the reconstructed phonon distribution. In \textbf{Fig.}\ref{fig:coherent displacement}(a), the coherent drive is off and we perform a least-squared fit on the qubit spectrum assuming thermal distribution, 
\begin{equation}
    P(n,n_\text{th}) = \frac{n_\text{th}^n}{(1+n_\text{th})^{n+1}},
\end{equation}
to find $n_\text{th}$ and the bare qubit frequency (See Sec.\ref{sec:map}). In \textbf{Fig.}\ref{fig:coherent displacement}(b-f), $n_{\text{disp}}$ is extracted by performing least-squared fits assuming displaced thermal distribution with $n_{\text{disp}}$ being the only free parameter. 

The coherent drive amplitude controls the phonon displacement, $n_{\text{disp}}\propto V_{\text{coh}}^2$. In \textbf{Fig.}\ref{fig:coherent displacement}(g), we verify the coherent displacement by plotting $V_{\text{coh}}^2$ against the extracted $n_{\text{disp}}$. The linear fit (dashed line) goes through the origin.
The $90\%$ confidence interval error bars on the extracted $n_{\text{disp}}$ can be found with non-parametric bootstrapping\cite{bootstrap,PRLsupplementary}: each qubit spectrum shown here is the average of 200 independently measured traces. We first create sets of synthetic data by re-sampling randomly with replacement among those 200 traces and then averaging. On those synthetic data, we perform the same displaced thermal fits to create a distribution of the extracted $n_{\text{disp}}$, from which a confidence interval can be found.

\section{ac-dither sideband} \label{sec:acDither}

We adopt the technique of driving ac-dither sidebands to access single phonon sideband transitions. Discussed in reference \cite{Blais2007Quantum-informationElectrodynamics} for a CPB qubit, this technique introduces a dynamical $\hat{\sigma}_z$ coupling by applying an ac-dither on the gate-charge, $n_g(t) = n_{g}^0 + n_g^{\text{dither}}\cos{(\omega_{\text{dither}}t)}$. With a small dither amplitude ($8E_c n_g^{\text{dither}} / E_J \approx 0.306$), the ac-dither sideband Hamiltonian is given by
\begin{equation}\label{eqn:dither Hamiltonian}
\begin{aligned}
    H_{BSB}^{\text{dither}} &= \Omega_{SB,0}\left(\hat{a}\hat{\sigma}_- + \hat{a}^\dagger \hat{\sigma}_+ \right),\\
    H_{RSB}^{\text{dither}} &= \Omega_{SB,0}\left(\hat{a}\hat{\sigma}_+ + \hat{a}^\dagger \hat{\sigma}_- \right),
\end{aligned}
\end{equation}
where
\begin{equation}\label{eqn:dither rate}
    \Omega_{SB,0} = g_m \frac{\Omega_R}{2\left(\omega_d - \omega_q \right)}J_1\left(\frac{8E_c n_g^{\text{dither}}}{E_J} \right)
\end{equation}
is the bare sideband rate, $\Omega_R$ and $\omega_d$ are the sideband drive Rabi rate and frequency, and $J_1(z)$ is the 1st order Bessel function of the first kind. The blue and red sideband transitions are located at frequencies
\begin{equation}\label{eqn:dither SB frequency}
    \begin{aligned}
        \omega_B(n) &= \omega_q^n + \omega_m \pm \omega_\text{dither},\\
        \omega_R(n) &= \omega_q^n - \omega_m \pm \omega_\text{dither}.
    \end{aligned}
\end{equation}

\begin{figure} [h!]
    \centering
    \includegraphics{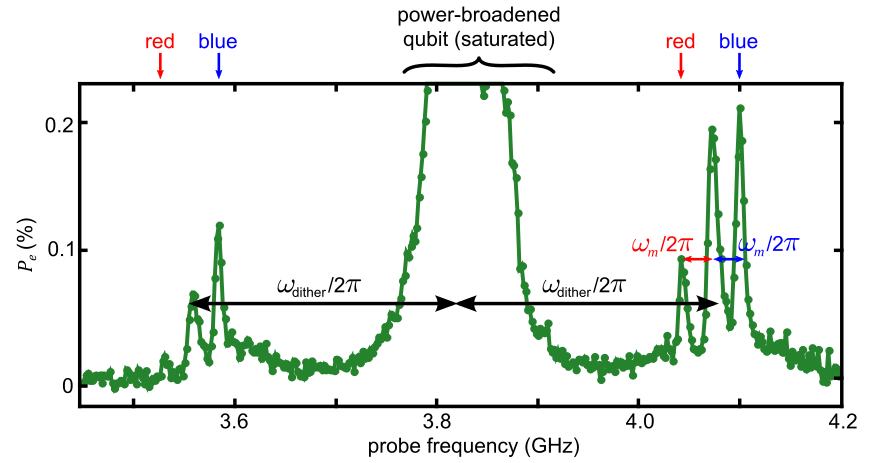}
    \caption{
    \textbf{Sideband spectrum at large spectroscopic power} \\\hspace{\textwidth}
    We measure the qubit spectrum at large probe power and $V_\text{dc} = 6$~V to observe the ac-dither sideband transitions. The lower (left) set of red and blue sideband transitions are used to manipulate the phonon populations in this work.
    }
    \label{fig:SBspectrum}
\end{figure}

Figure \ref{fig:SBspectrum} shows a full measured qubit spectrum under an ac-dither at frequency $\omega_\text{dither} = 2\pi \times260$~MHz with a large spectroscopic power. The center qubit transition is power broadened\cite{Schuster2005AcField} and saturated, with $P_e \approx 0.5$ on resonance. Two groups of 3 satellite peaks are visible on either side of the main qubit peak, detuned by the ac-dither frequency. Within each group, we can identify the single phonon red and blue sideband transitions. For sideband related experiments described in the main text, we apply an ac-dither drive at frequency $\omega_\text{dither} = 2\pi \times257$~MHz.

\section{Modeling the dynamics of the system}\label{sec:dynamics}

To understand and predict the dynamics of the qubit-mechanics system, we employ a classical master-equation calculation. In Sec.\ref{sec:master eqn}, we outline the theory describing the system dynamics. Particularly, instead of the simplified picture of Fig.3(c), we provide an accurate description of the decay process where, most notably, a qubit decay also connects different mechanical states. Using this theory, in Sec.\ref{sec:Fitting}, we extract the sideband driving parameters from measurements to produce Fig.3(f,i) in the main text.

\subsection{Classical master equation}\label{sec:master eqn}

We make two approximations to understand the system dynamics. First, we consider the qubit-mechanics system to be fully described by the diagonal elements of its density matrix, $P_g^{\ket{n}}$ and $P_e^{\ket{n}}$. Second, we approximate the system evolution under sideband drives by a classical master equation similar to the classical laser rate equations. Those approximations are valid because we drive sidebands on a time scale much longer than the qubit dephasing time ($T_2^* \approx 80$~ns) and relaxation time ($T_1 \approx 260$~ns) and with a rate $\Omega_{SB,0} \lesssim 2\pi \times 100$~kHz (see Sec.\ref{sec:Fitting}) much smaller than qubit decoherence rate $\Gamma_2^* \approx 2 \pi \times 3.7$~MHz.

The overall master equation is given by a combination of system dynamics due to sideband drives and decay,
\begin{equation}\label{eqn:master eqn overall}
    \frac{d}{dt}P_{g(e)}^{\ket{n}} = \frac{d}{dt}P_{g(e)}^{\ket{n}}\bigg \vert_{\text{drive}} + \frac{d}{dt}P_{g(e)}^{\ket{n}}\bigg \vert_{\text{decay}}.
\end{equation}

To start, we first expand the first term in eqn.(\ref{eqn:master eqn overall}), which describes the dynamics due to sideband driving\cite{PRLsupplementary},
\begin{equation}\label{eqn:master eqn drive}
    \begin{aligned}
        \frac{d}{dt}P_g^{\ket{n}}\bigg \vert_{\text{drive}} &= \begin{aligned}[t]
            -\bigg( \Gamma_B^{\ket{n}}(n_B) &+\Gamma_R^{\ket{n}}(n_R) \bigg) P_g^{\ket{n}} \\ &+ \Gamma_R^{\ket{n}}(n_R)P_e^{\ket{n-1}}+\Gamma_B^{\ket{n}}(n_B)P_e^{\ket{n+1}},
        \end{aligned}\\
        \frac{d}{dt}P_e^{\ket{n}}\bigg \vert_{\text{drive}} &= \begin{aligned}[t]
             -\bigg( \Gamma_B^{\ket{n}}(n_B) &+\Gamma_R^{\ket{n}}(n_R)\bigg) P_e^{\ket{n}} \\ &+ \Gamma_R^{\ket{n}}(n_R)P_e^{\ket{n+1}}+\Gamma_B^{\ket{n}}(n_B)P_e^{\ket{n-1}},
        \end{aligned}
    \end{aligned}
\end{equation}
where $P_{g(e)}^{\ket{n}}$ is the instantaneous population in the qubit ground (excited) state with $n$ phonons. $\Gamma_B^{\ket{n}}(n_B)$ is the reduced blue sideband rate of transition $\ket{g,n}\leftrightarrow \ket{e,n+1}$ when the blue sideband drive is resonant with transition $\ket{g,n_{B}}\leftrightarrow \ket{e,n_{B}+1}$,
\begin{equation}\label{eqn:reduced dither blue rate}
    \Gamma_B^{\ket{n}}(n_B) = \frac{4(n+1)\Omega_{BSB,0}^2}{\Gamma_2^*}\frac{1}{1+\left( \frac{4\chi_m(n-n_B)}{\Gamma_2^*} \right)^2}.
\end{equation}
When the blue sideband drive is chirped, $n_B$ is a function of time.
Conversely, $\Gamma_R^{\ket{n}}(n_R)$ is the reduced red sideband rate of transition $\ket{g,n}\leftrightarrow \ket{e,n-1}$ when the red sideband drive is resonant with transition $\ket{g,n_{R}}\leftrightarrow \ket{e,n_{R}-1}$,
\begin{equation}\label{eqn:reduced dither red rate}
    \Gamma_R^{\ket{n}}(n_R) = \frac{4n\Omega_{RSB,0}^2}{\Gamma_2^*}\frac{1}{1+\left( \frac{4\chi_m(n-n_R)}{\Gamma_2^*} \right)^2}.
\end{equation}

Next, we write down the master equations due to qubit and mechanical decay. The ultra-strong coupling introduces new decay channels to the system, and complicates the simplified picture of Fig.3(c) in the main text. 
Because $\hbar \omega_m \ll k_B T \ll \hbar \omega_q$, we allow for thermal excitations in the mechanical oscillator, but assume no thermal population in the qubit. Following reference \cite{Beaudoin2011DissipationQED}, we find these master equations to be,
\begin{equation}\label{eqn:master eqn decay USC}
    \begin{aligned}
    \frac{d}{dt}P_g^{\ket{n}}\bigg \vert_{\text{decay}} &=  
        \begin{aligned}[t]
            e^{2r} \gamma_m \bigg\{ n_\text{th}n P_g^{\ket{n-1}} + &(1+n_\text{th})(1+n)P_g^{\ket{n+1}} - \bigg[(1+n)n_\text{th} + (1+n_\text{th})n \bigg]P_g^{\ket{n}} \bigg\}\\
            &+ \Gamma_1 \sum_m \abs{\alpha_{mn}}^2 P_e^{\ket{m}}
        \end{aligned}\\
    \frac{d}{dt}P_e^{\ket{n}}\bigg \vert_{\text{decay}} &=
        \begin{aligned}[t]
            e^{-2r} \gamma_m \bigg\{ n_\text{th}n P_e^{\ket{n-1}} + &(1+n_\text{th})(1+n)P_e^{\ket{n+1}} - \bigg[(1+n)n_\text{th} + (1+n_\text{th})n \bigg]P_e^{\ket{n}} \bigg\}\\
            &- \Gamma_1 P_e^{\ket{n}}
        \end{aligned}\\
    \end{aligned},
\end{equation}
where $r = \chi_m / 2 \omega_m$ is the squeezing amplitude, and $\alpha_{mn} = \prescript{}{b}{\bra{m}}\hat{S}(-2r)\ket{n}_b$ is the overlap between different motional states under a qubit flip (Sec.\ref{sec:USC}). Here, we ignore the relaxation process where the mechanical oscillator losses many phonons to excite the qubit ($\ket{g,n}\mapsto \ket{e,m}$). 
Because of the large difference between qubit and mechanical frequency, such process requires the loss of more than 152 phonons ($\omega_q / \omega_m \approx 152$), and is highly unlikely.

\begin{figure} [h!]
    \centering
    \includegraphics{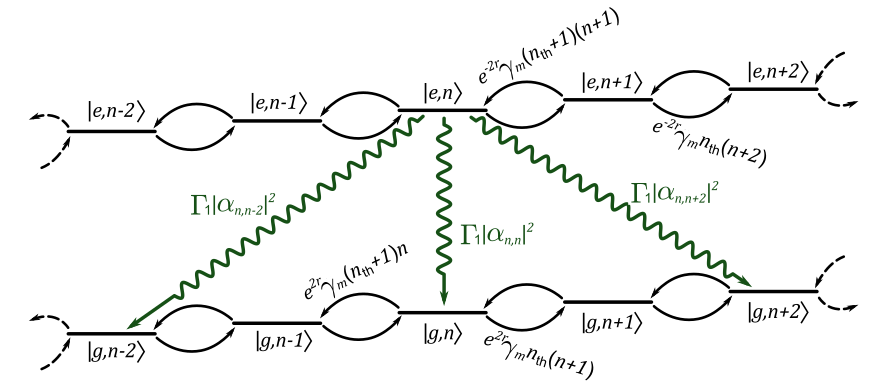}
    \caption{
    \textbf{System decay under ultra-strong coupling}\\
    Compared to Fig.3(c) of the main text, the decay dynamics of the qubit-mechanics system have two main differences. 
    First, the qubit decay is no longer restricted to be phonon number conserving. Instead, a decaying qubit can cause changes in the phonon population, analogous to sideband transitions. 
    Second, the mechanical decay rate gains a scaling factor that is qubit-state dependent. Because $2r \approx 0.01 \ll 1$, this factor is very close to unity.}
    \label{fig:USCDecay}
\end{figure}

As illustrated in \textbf{Fig.}\ref{fig:USCDecay}, the master equations of eqn.(\ref{eqn:master eqn decay USC}) contain two corrections compared to the simplified picture given in Fig.3(a) of the main text. First, a qubit decay can connect different mechanical states. As discussed in Sec.\ref{sec:FrankCondon}, this corresponds to a fast electronic decay connecting different mechanical states. Similar to the sideband transitions, these phonon non-conserving qubit decays become dominant at large phonon number, and strongly affect the dynamics of the qubit-mechanics system. Second, the mechanical decay rate is qubit-state dependent ($e^{-2r \hat{\sigma}_z}\gamma_m$). For us, this is a very small correction because $2r = \chi_m/\omega_m \approx 0.01$. 

To perform numerical simulations described in this paper, we truncate the mechanical phonon space at $n_\text{max}=200$, and numerically solve the set of differential equations in eqn.(\ref{eqn:master eqn overall}). At every time step in the simulation, we enforce the conservation of probability $\sum_{n=0}^{n_\text{max}} \left(P_g^{\ket{n}} + P_e^{\ket{n}}\right) = 1$ by setting $P_e^{\ket{n_\text{max}}} = 0$ and $P_g^{\ket{n_\text{max}}} = 1-\sum_{n=0}^{n_\text{max}-1} \left(P_g^{\ket{n}} + P_e^{\ket{n}} \right)$. 

\subsection{Extracting the sideband driving parameters (Fig.3(f,i) of the main text)}\label{sec:Fitting}

In this section, we describe the process of extracting the parameters of sideband drives to generate the simulations shown in Fig.3(f,i).
We extract those parameters by performing least-squared fits to the measured qubit spectra corresponding to Fig.3(e,h). Specifically:

(1) We find $n_\text{th}$ by fitting the $\tau = 0$ data with thermal distribution, where $\tau$ is the chirp time. 

(2) We generate a look-up table of phonon distributions with different sideband driving parameters by numerically solving eqn.(\ref{eqn:master eqn overall}). Through a convolution with the PSF map (Sec.\ref{sec:map}), this table is converted into a look-up table of expected qubit spectra.

\begin{figure} [h!]
    \centering
    \includegraphics{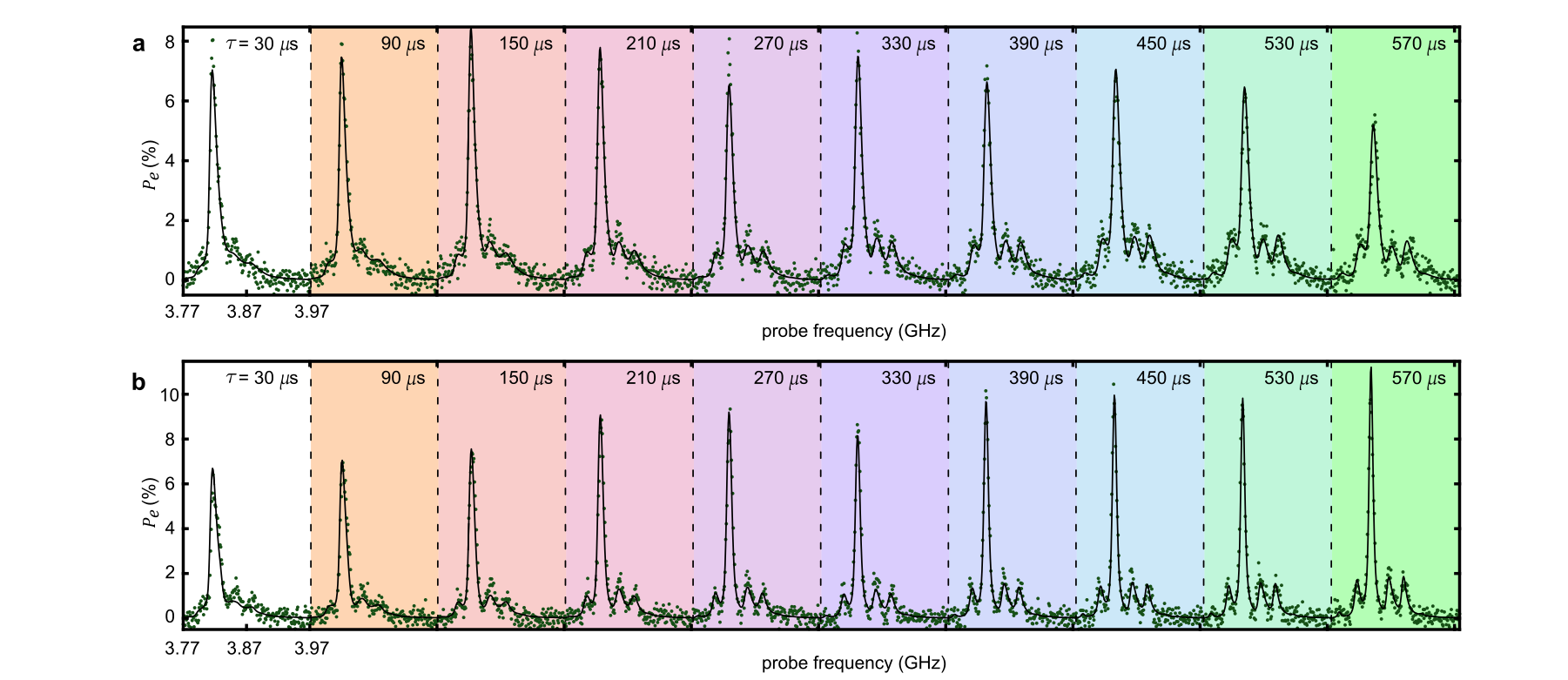}
    \caption{
    \textbf{Extracting the sideband driving parameters}\\
    \textbf{a,} With the red sideband drive off, we chirp the blue sideband drive for different time $\tau$ and measure the qubit spectrum. All panels have the same frequency-axis. Using a least-squared fit on the thermal state qubit spectrum at $\tau = 0$, we find a mechanical thermal population of $n_\text{th} = 13$. We experimentally determine the blue sideband chirp rate to be 38.5 phonons/ms.
    Using a joint least-squared fit to the look-up table generated by convolving the numerical solutions of eqn.(\ref{eqn:master eqn overall}) with the qubit response to phonon Fock states, we find the only free parameters: $n_B(0) = -1.5$ and $\Omega_{BSB,0} = 2\pi \times 89$~kHz.  
    \textbf{b,} Keeping the same blue sideband drive parameters, we turn on the red sideband drive and repeat the measurements in \textbf{a}. We determine $n_R = 44$ using the detuning between the sideband drive frequencies and $n_B(0)$. A least-squared fit to the thermal state finds $n_\text{th} = 15$. Finally, a joint least-squared fit to the look-up table finds the only free parameter $\Omega_{RSB,0} = 2\pi \times 66$~kHz.}
    \label{fig:joint fit}
\end{figure}

(3) Using the look-up table, we jointly fit all data with only a blue sideband drive but chirped for different $\tau$. Because the chirp rate is determined experimentally, the only two free parameters are the bare blue sideband rate $\Omega_{BSB,0} = 2\pi \times 89$~kHz, and the starting position of the chirp $n_B(0) = -1.3$. (\textbf{Fig.}\ref{fig:joint fit}(a))

(4) Finally, we jointly fit all data with both blue and red sideband drives. Because we use the same blue sideband drive protocol, we can keep the fit results from step(3) as constants. The center of transition for the red sideband drive $n_R = 44$ is determined through eqn.(\ref{eqn:dither SB frequency}) using the fit result of $n_B(0)$ and the frequency detuning between the initial blue and red sideband drives. A joint least-squared fit to the look-up table finds the only free parameter $\Omega_{RSB,0} = 2\pi \times 66$~kHz.  (\textbf{Fig.}\ref{fig:joint fit}(b))
  
\section{Limitation on Fano factors from using ac-dither sideband}\label{sec:Limit}

\begin{figure} [h!]
    \centering
    \includegraphics{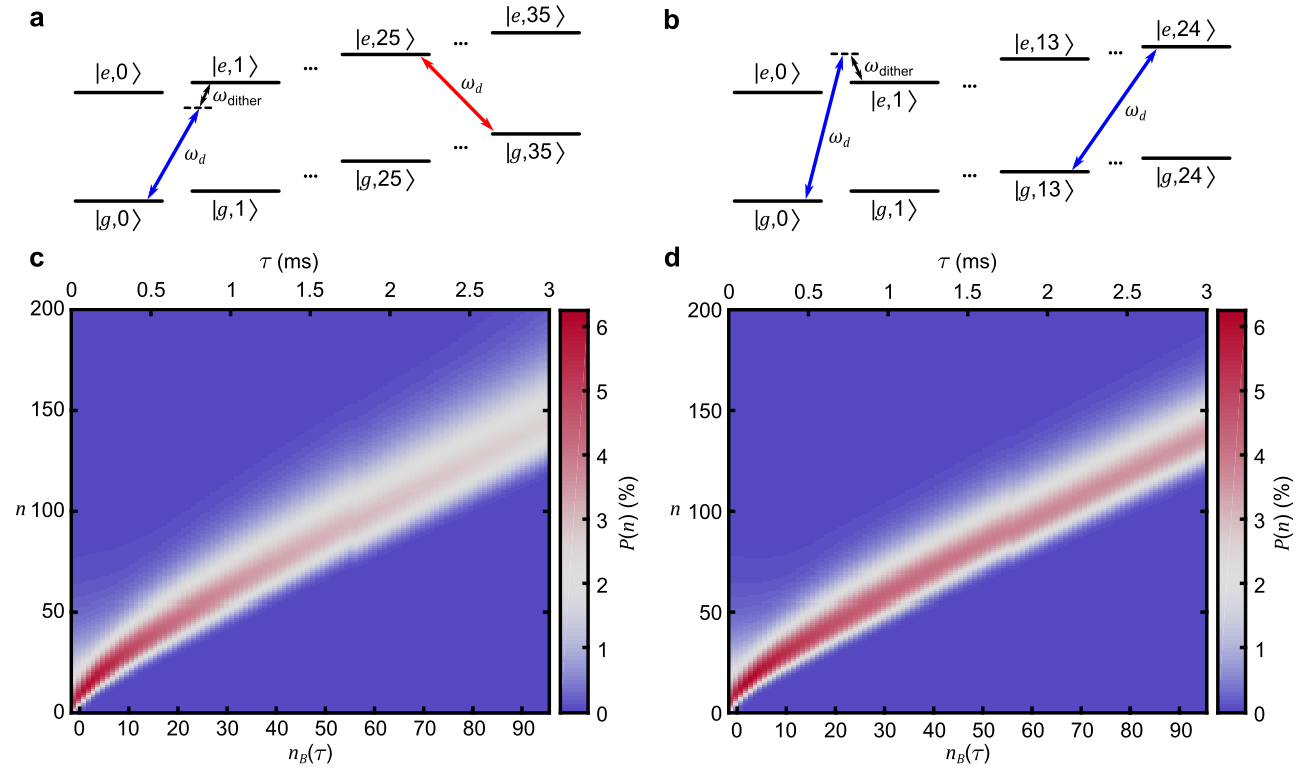}
    \caption{
    \textbf{Spurious cooling and heating}\\
    \textbf{a,} A sideband drive at frequency $\omega_d = \omega_q^0-\omega_\text{dither}+\omega_m$ is simultaneously resonant with the ac-dither blue sideband transition of $\ket{g,0}\leftrightarrow \ket{e,1}$ and the 10-th order red sideband transition of $\ket{g,35}\leftrightarrow \ket{e,25}$. Thus, using this ac-dither blue sideband drive to add phonons to the mechanical oscillator will cause unwanted cooling at larger phonon number.
    \textbf{b,} Similar to \textbf{a}, a blue sideband drive at frequency $\omega_d = \omega_q^0+\omega_\text{dither}+\omega_m$ is simultaneously resonant with the ac-dither blue sideband transition of $\ket{g,0}\leftrightarrow \ket{e,1}$ and the 11-th order blue sideband transition of $\ket{g,13}\leftrightarrow \ket{e,24}$. This will cause unwanted heating.
    \textbf{c, d,} We use the master equations in Sec.\ref{sec:dynamics} and the driving parameters extracted in Sec.\ref{sec:Fitting} to calculate the phonon distribution (color-scale vs. y-axis) after chirping the ac-dither blue sideband drive for time $\tau$ (top-axis). This process should empty all phonon populations below the center of the blue sideband drive $n_B(\tau)$ (bottom-axis). The spurious cooling process is not considered in \textbf{c}, but it is modeled in \textbf{d}. The unwanted cooling is visible at $n_B(\tau) > 50$.
    }
    \label{fig:spuriousSqueezing}
\end{figure}

We demonstrate the sub-Poissonian state preparation at a moderate mean phonon number $\langle n \rangle = 43$ because driving the ac-dither sideband introduces an additional unwanted process that significantly alters the phonon population at large phonon number $n > 50$. Specifically, when a blue sideband drive is chirped for time $\tau$ to position $n_B(\tau)$ to empty all phonon populations below, it simultaneously cools the populations at larger phonon numbers. This additional unwanted cooling process (spurious cooling) is conceptually illustrated in \textbf{Fig.}\ref{fig:spuriousSqueezing}(a). We use the set of ac-dither sidebands below the qubit resonance, $\omega_{B(R)}(n) = \omega_q^n \pm \omega_m - \omega_\text{dither}$ (\textbf{Fig.\ref{fig:SBspectrum}}). Meanwhile, the $l$-th order sideband transition is located at $\omega_l(n) = \omega_q^n + l\omega_m$. Thus, an ac-dither blue sideband drive centered on transition $\ket{g,n}\leftrightarrow \ket{e,n+1}$ can be close to resonance with the 10-th order red sideband transition $\ket{g,n+35}\leftrightarrow\ket{e,n+25}$, creating an additional cooling effect at larger phonon numbers.

In addition to the resonant 10-th order red sideband transition, an ac-dither blue sideband drive can also excite lower order red sideband transitions off-resonantly, which dominates this spurious cooling process. To understand this, we write the reduced $l$-th order sideband rate for transition $\ket{g,n}\leftrightarrow\ket{e,n+l}$ due to an ac-dither sideband drive at $\omega_d$ (see Sec.\ref{sec:master eqn}),
\begin{equation}\label{eqn:spurious cooling lth order reduced rate}
\begin{aligned}
    \Gamma_l^{\ket{n}}(\omega_d) &= \frac{4 \abs{\Omega_n^l(\omega_d)}^2}{\Gamma_2^*} \frac{1}{1+\left( \frac{ \omega_d - \omega_l(n)}{\Gamma_2^*/2} \right)^2}\\
    & = \frac{4\Omega_R^2(\omega_d)}{\Gamma_2^*} {\abs{\alpha_{n+l,n} ^\text{rsd}}}^2 \frac{1}{1+\left( \frac{\omega_d - \omega_l(n)}{\Gamma_2^*/2} \right)^2},
\end{aligned}
\end{equation}
where $\alpha_{n+l,n} ^\text{rsd}$ is the overlap given by eqn.(\ref{eqn:mech state overlap sigma_z}). At smaller $l$, the increased overlap $\abs{\alpha_{n+l,n} ^\text{rsd}}^2$ more than makes up the larger detuning. Moreover, the larger detuning at smaller $l$ also creates a more uniform sideband rate $\Gamma_l^{\ket{n}}(\omega_d)$ as a function of $n$, resulting in a more efficient manipulation of phonons\cite{PRLsupplementary}.

In \textbf{Fig.}\ref{fig:spuriousSqueezing}(d), we numerically demonstrate the effect of this spurious cooling process. 
We use the calculated qubit spectrum of \textbf{Fig.}\ref{fig:classical vs. quantum map}(c) to find $\abs{\alpha_{n+l,n} ^\text{rsd}}^2$. With the driving parameters extracted from Sec.\ref{sec:Fitting},  we incorporate up to the 10-th order red sideband transitions into the master equation (Sec.\ref{sec:master eqn}) to predict the phonon distribution under a chirped ac-dither blue sideband drive. 
Compared to \textbf{Fig.}\ref{fig:spuriousSqueezing}(c) where the red sideband transitions are ignored, the effect of the spurious cooling is clear. This result is consistent with what we observe experimentally. 

Although this spurious cooling process helps in achieving a narrow phonon distribution, we instead operated at a relatively small phonon number where this effect is negligible. This allows us to achieve better control over the qubit-mechanics system. Note, in the joint fits described in Sec.\ref{sec:Fitting}, the spurious cooling is not modeled.

Conversely when driving the ac-dither sideband above the qubit, we also observe spurious heating. This is conceptually illustrated in \textbf{Fig.}\ref{fig:spuriousSqueezing}(b), and a similar discussion as above can be conducted. 

\section{Bounding the true Fano factor}\label{sec:Bound}

 We repeatedly perform the reconstruction procedures on simulated experiments to understand the statistical significance of the extracted Fano factor, and consequently place a bound on the true Fano factor of the sub-Poissonian state that we prepare in the main text. This process is necessary because the extracted phonon distribution need not be a faithful reconstruction of the true mechanical occupation, but could be biased by individual measurement noise realizations or reconstruction procedures. 
 
 The four step procedure to conduct this process is as following: 
 
 (1) We create a Gaussian distributed phonon distribution characterized by a mean phonon number $\langle n \rangle = 43$ and a Fano factor $F_\text{true}$. 
 
 (2) This phonon distribution is converted into a noiseless qubit spectrum by convolving it with the PSF map found in Sec.\ref{sec:map}. 
 
 (3) We simulate an experimental realization by adding the appropriate amount of noise and averaging for the same number of times as the actual experiment (916 times for the data in Fig.4 of main text). 
 
 (4) The same reconstruction procedure is performed on the simulated experiment to extract a phonon distribution and $F_\text{extract}$. 
 
\begin{figure} [h!]
    \centering
    \includegraphics{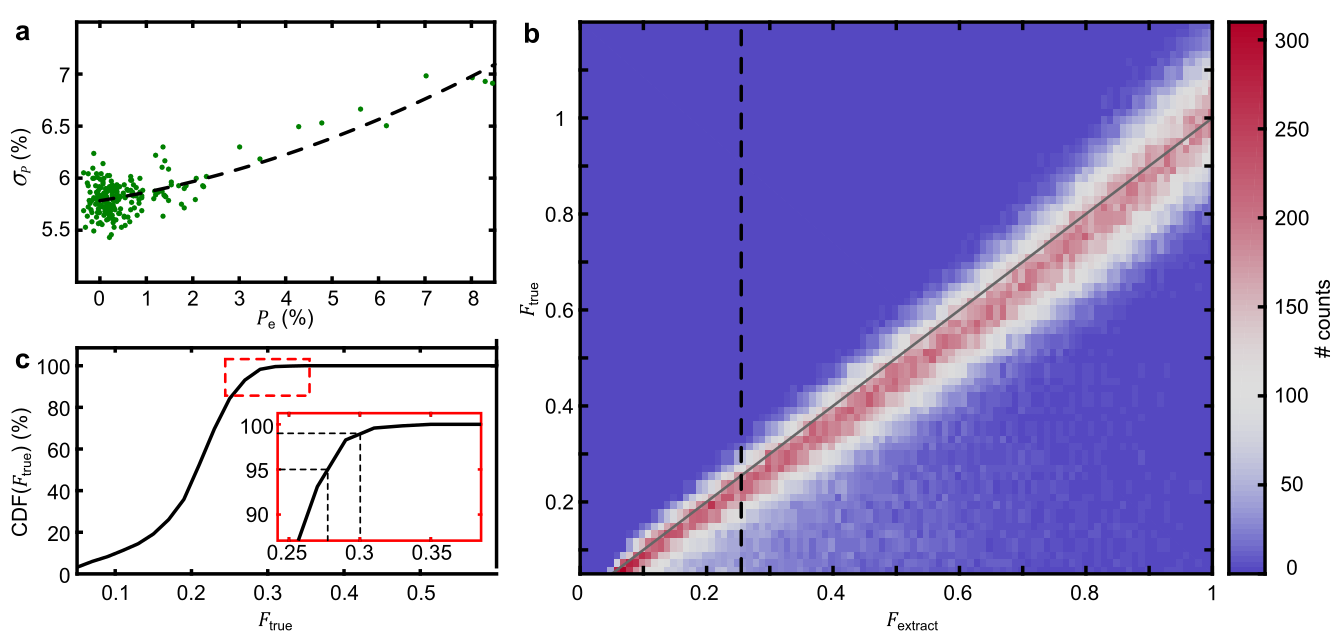}
    \caption{
    \textbf{Bounding the Fano factor}\\
    \textbf{a,} We plot the measurement noise $\sigma_P$ as a function of the mean measured signal. The dashed black line is a second order polynomial fit, which we use to model the noise dependence on signal amplitude.
    \textbf{b,} We plot the histogram of $F_\text{extract}$ (color-scale vs. x-axis) for different $F_\text{true}$. Each $F_\text{true}$ consists of 3000 different simulated experiments. The solid grey line corresponds to $F_\text{extract}=F_\text{true}$, and shows that $F_\text{extract}$ in general overestimates the Fano factor.
    \textbf{c,} Using the CDF of $F_\text{true}$ for $F_\text{extract} \in [0.255,0.256]$ (dashed black line in \textbf{b}), we find with 95\% confidence that $F_\text{true} \leqslant 0.28$ and with 99\% confidence that $F_\text{true} \leqslant 0.30$.}
    \label{fig:BoundingFanoFactor}
\end{figure}

Correctly simulating the experiment in step(3) requires an accurate understanding of the measurement noise. We achieve this understanding by looking at the measurement data that resulted in Fig.4 of the main text. We can find the standard deviation of the measurement signal at each frequency point because we have recorded all 916 individual qubit traces. At each frequency point, the extracted measurement standard deviation $\sigma_P$ (green dot) is plotted in \textbf{Fig.}\ref{fig:BoundingFanoFactor}(a), and arranged according to the mean signal at that frequency. We model the dependence of noise on signal amplitude with a second order polynomial function (black dashed line).

The four-step procedure is repeated many times for different noise realizations and $F_\text{true}$ to build statistics. Figure \ref{fig:BoundingFanoFactor}(b) depicts the result of this statistical study. For each $F_\text{true}$, we perform 3000 simulated experiments, each with 916 averages. The extracted $F_\text{extract}$ are binned into steps of $\delta F = 0.01$, and the resulting histogram is plotted. The grey solid line corresponds to $F_\text{extract} = F_\text{true}$, and demonstrates that $F_\text{extract}$ in general overestimates the Fano factor. For the 1216 simulations that resulted in an $F_\text{extract}$ within the interval $[0.255,0.265]$ (black dashed line), the cumulative distribution function (CDF) of $F_\text{true}$ is plotted in \textbf{Fig.}\ref{fig:BoundingFanoFactor}(c). Because the experimentally determined $F = 0.257^{+0.002}_{-0.001}$ lies within this range, we thus find $F_\text{true} \leqslant 0.28$ with 95\% confidence and $F_\text{true} \leqslant 0.30$ with 99\% confidence for the sub-Poissonian state we prepared in the main text.

\section{Negativity in the Wigner function}

Because any state prepared using our energy-squeezing protocol will contain only diagonal elements, we can easily compute the Wigner function from the extracted phonon distribution. Doing so with the sub-Poissonian state in Fig.4 of the main text, we observe a small region of negativity in the Wigner function but with high statistical significance. Unfortunately, the qubit frequency is a kind of nuisance parameter whose uncertainty yields systematic uncertainty in the average phonon number $\langle n \rangle$. Although this uncertainty has little effect on the Fano factor, it causes uncertainty in the location of the Wigner negativity. If we average over all of the possible Wigner functions consistent with the possible qubit frequencies, the negativity is diminished such that it has marginal statistical significance. 

\begin{figure} [h]
    \centering
    \includegraphics{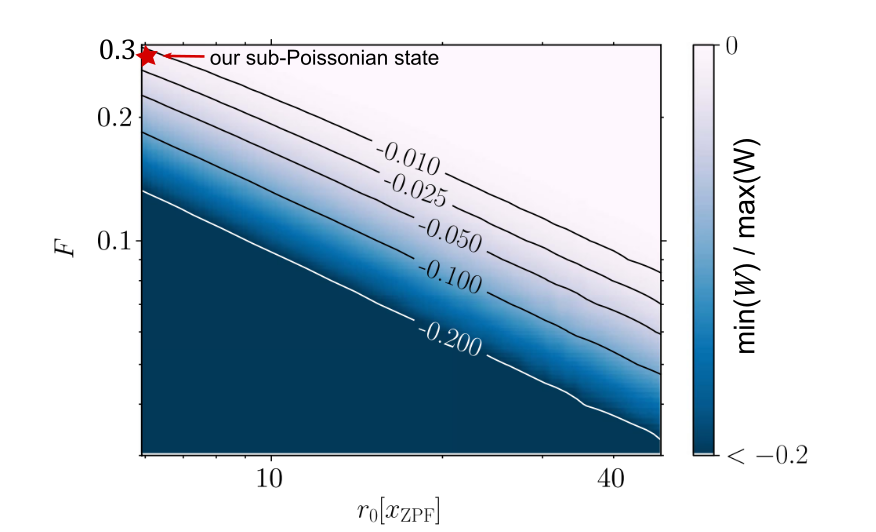}
    \caption{
    \textbf{Fano factor and negativity in the Wigner function}\\
    For a Gaussian-distributed sub-Poissonian state, the maximal negativity in the Wigner function (color-axis) can be related to its motional amplitude $r_0$ (x-axis) and Fano factor (y-axis). In comparison, assuming a Gaussian number fluctuation, the sub-Poissonian state we prepare in Fig.4 ($r_0 = \sqrt{43}$ and $F \leq 0.28$ with 95\% confidence) is located in this figure at the red-stared position, indicating a small negativity in its Wigner function. This figure by L\"orch \textit{et al.}\cite{Lorch2014LaserRegime} is reproduced under \href{https://creativecommons.org/licenses/by/3.0/}{CC BY 3.0} license. We added the red star, its associated labels, and the color-axis label ``$\text{min}(W)/\text{max}(W)$''.
    }
    \label{fig:Wigner}
\end{figure}

Instead of further analyzing the data and risk introducing bias, we can relate the extracted Fano factor to a negativity in the Wigner function under the assumption of a Gaussian number fluctuations, according to reference\cite{Lorch2014LaserRegime}. 
In this reference, the authors finds that for a Gaussian-distributed sub-Poissonian state, the maximal negativity in the Wigner $W$ function, defined as the ratio of $\text{min}(W)/\text{max}(W)$, is related to its motional amplitude $r_0 = \sqrt{\langle n \rangle}$ and Fano factor $F$ according to Fig.6 of the reference, reproduced in Fig.~\ref{fig:Wigner}. In comparison, the sub-Poissonian state we prepare in this work is approximately Gaussian-distributed around its mean phonon number $\langle n \rangle = 43$, and has a Fano factor $F \leq 0.28$ with 95\% confidence. Under the assumption of a Gaussian number fluctuations, these values place us at the red-stared position in this figure, and indicate that our nonclassical state possesses a small negativity in the Wigner function.

\newpage

\bibliographystyle{naturemag} 
\bibliography{SupplementBib.bib}